\documentclass[review]{elsarticle}

\usepackage{hyperref}
\usepackage{enumitem}
\usepackage{subfig}
\usepackage{longtable}
\usepackage{amsmath}
\usepackage{graphicx}

\journal{Information Fusion}

\begin{document}

\begin{frontmatter}

\title{CochleaNet: A Robust Language-independent Audio-Visual Model for Speech Enhancement}

\author[mymainaddress]{Mandar Gogate}
\author[stir]{Kia Dashtipour}
\author[mysecondaryaddress]{Ahsan Adeel}
\author[mymainaddress]{Amir Hussain\corref{mycorrespondingauthor}}

\cortext[mycorrespondingauthor]{Corresponding author}
\ead{a.hussain@napier.ac.uk}

\address[mymainaddress]{Edinburgh Napier University, School of Computing, Edinburgh, EH10 5DT, UK}
\address[stir]{University of Stirling, Division of Computing Science and Maths, Stirling, FK9 4LA, UK}
\address[mysecondaryaddress]{University of Wolverhampton, School of Mathematics and Computer Science, Wolverhampton, UK}

\begin{abstract}
Noisy situations cause huge problems for suffers of hearing loss as hearing aids often make the signal more audible but do not always restore the intelligibility.
In noisy settings, humans routinely exploit the audio-visual (AV) nature of the speech to selectively suppress the background noise and to focus on the target speaker.
In this paper, we present a causal, language, noise and speaker independent AV deep neural network (DNN) architecture for speech enhancement (SE).
The model exploits the noisy acoustic cues and noise robust visual cues to focus on the desired speaker and improve the speech intelligibility.
To evaluate the proposed SE framework a first of its kind AV binaural speech corpus, called ASPIRE, is recorded in real noisy environments including cafeteria and restaurant.
We demonstrate superior performance of our approach in terms of objective measures and subjective listening tests over the state-of-the-art SE approaches as well as recent DNN based SE models.
In addition, our work challenges a popular belief that a scarcity of multi-language large vocabulary AV corpus and wide variety of noises is a major bottleneck to build a robust language, speaker and noise independent SE systems.
We show that a model trained on synthetic mixture of Grid corpus (with 33 speakers and a small English vocabulary) and ChiME 3 Noises (consisting of only bus, pedestrian, cafeteria, and street noises) generalise well not only on large vocabulary corpora but also on completely unrelated languages (such as Mandarin), wide variety of speakers and noises.
\end{abstract}

\begin{keyword}
\texttt{Audio-Visual\sep Speech Enhancement\sep Speech Separation \sep Deep Learning \sep Real Noisy Audio-Visual Corpus \sep Speaker Independent \sep Causal}
\end{keyword}

\end{frontmatter}

\section{Introduction}\label{sec:introduction}
The human auditory cortex has a remarkable capability to focus on a target speech by selectively suppressing the ambient noise. The selective suppression of unwanted background noise is known to exploit the noise robust visual cues to enhance a person's capacity to resolve the phonological ambiguities~\cite{golumbic2013visual}.
In addition, studies have shown the importance of visual cues in improving the speech intelligibility~\cite{summerfield1992lipreading} as well as speech detection in noisy environments~\cite{grant2000use,grant2001speech}.
In this study, we achieve this selective speech enhancing ability computationally.

In the recent years, speech enhancement (SE) has attracted wide attention due to the noise reducing ability that helps hearing impaired listen better in noisy social situations and opened the doors for speech processing systems (such as speech recognition and voice activity detector systems) in noisy environments~\cite{narayanan2014investigation, kayser2015improving}.
SE approaches can be categorised into statistical analysis based noise reduction models (such as spectral subtraction (SS), linear minimum mean square error  (LMMSE) and Wiener filtering) and computational auditory scene analysis (CASA)~\cite{wang2006fundamentals}.
It has been observed that, the statistical methods fail to achieve improved speech intelligibility in some scenarios due to introduction of distortions such as musical noises.
In contrast, CASA has shown to be more effective in stationary and non-stationary noises~\cite{chen2018dnn}.

In CASA, the speech is separated from interfering background noise by using a time-frequency (T-F) spectral mask to the T-F representation of noisy speech. The T-F spectral mask is used to enhance speech dominant regions and suppress the noise-dominant regions. The ideal binary mask (IBM) assigns zero to a T-F unit if the local signal-to-noise ratio (SNR) is lower than the local criterion (LC), and unit value otherwise. The IBM is defined as follows:

\begin{equation}
IBM(t,f) =
     \begin{cases}
       \text{0} &\quad\text{if $SNR (t,f) \leq$ LC}\\
       \text{1} &\quad\text{$otherwise$.} \\
     \end{cases}
\label{equation:IBM}
\end{equation}

The IBM has shown to improve the speech quality and intelligibility for the hearing impaired and normal hearing listeners~\cite{kjems2010speech, ahmadi2013perceptual, wang2009speech}.
The IBM cannot be calculated using equation~\ref{equation:IBM} in real-world scenarios because the target speech and interfering background noise cannot be estimated with high accuracy.
However, the IBM estimation can be modelled as a data-driven optimisation problem that jointly exploits noisy speech and visual face images for the spectral mask estimation.

\begin{figure}[!t]
  \centering
  \includegraphics[width=\linewidth]{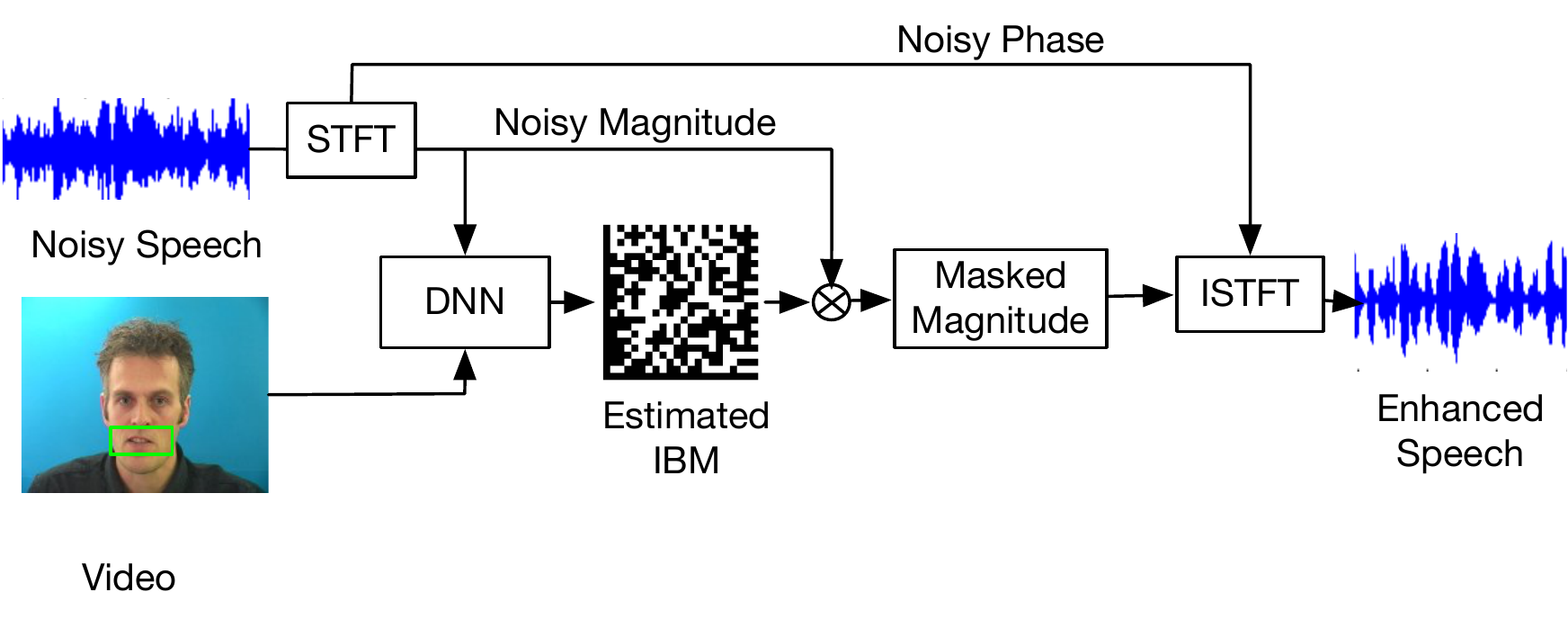}
  \caption{CochleaNet Framework: Audio-Visual Mask Estimation based Speech Enhancement}
  \label{fig:overview}
\end{figure}

In the literature, extensive research has been carried out to develop audio-only (A-only) and audio-visual (AV) SE methods.
Researchers have proposed several SE models such as deep neural network (DNN) based spectral mask estimation models~\cite{ephrat2018looking, gogate2018dnn}, DNN based clean spectrogram estimation models~\cite{gabbay2018visual, hou2018audio}, Wiener filtering based hybrid models~\cite{adeel2017towards, adeel2018lip, adeel2018contextual}, and time-domain SE models~\cite{rethage2018wavenet, pandey2018new, luo2019conv}.
However, limited work has been conducted to develop robust language-independent, causal, speaker and noise-independent AV SE models for low SNRs ($< - 3$  dB) observed in everyday social environments (such as cafeteria, and restaurants) where traditional A-only hearing aids fail to improve the speech intelligibility.
The few attempts to develop such robust models have been limited to speaker-dependent scenarios~\cite{hou2018audio} and small scale ($< 5 $ speakers) speaker independent scenarios~\cite{gogate2018dnn, adeel2018contextual}.

In addition, none of the aforementioned AV SE studies have conducted listening tests on real noisy mixtures that often consists of speech signal reverberantly mixed with multiple competing background noise sources~\cite{CHiME3}.
Finally, studies have shown that a pretrained DNN based SE model does not generalise well on new languages~\cite{pascual2018language}.
The model can be fine-tuned on large AV corpus consisting of wide variety of languages such as AVSPEECH~\cite{ephrat2018looking} (consisting of 1500 hours recording) to potentially achieve the language-independent performance given enough model capacity.
However, training on corpora like AVSPEECH requires a large number of graphics processing units (GPUs) or tensor processing units (TPUs) that are often unavailable in academic research environments.

In this paper, we present a causal, language, noise and speaker independent AV model to focus on a target speaker by selectively suppressing the background noise.
More specifically, we design and train a cross-modal DNN architecture, called CochleaNet, that ingests the noisy sound mixture and cropped images of speakers lip as an input and output a T-F mask to selectively suppress and enhance each T-F bin.
In addition, the model contextually exploits the available AV cues to estimate the spectral mask independent of the SNRs.

The proposed AV SE model is evaluated using, ASPIRE, a first of its kind high quality AV binaural speech corpus recorded in real noisy settings such as cafeteria and restaurant.
It is to be noted that, most of the aforementioned AV SE methods used a synthetic mixture of clean speech and noises for model evaluation.
However, the synthetic mixture do not reflect the real noisy mixtures as speech is often reverberantly mixed with multiple competing noise background sources.
Therefore, the ASPIRE corpus can be used by speech and machine learning communities as a benchmark resource to support reliable evaluation of AV SE technologies.

We demonstrate superior speech quality and intelligibility of proposed approach over the state-of-the-art A-only SE approaches as well as recent DNN based SE models.
In addition, we show that a model trained on a synthetic mixture of Grid corpus~\cite{Grid} (with only 33 speakers and a small English vocabulary) and ChiME 3~\cite{CHiME3} noises (consisting of bus, pedestrian, cafe, and street noises) generalise well on real noisy ASPIRE corpus, large vocabulary corpora (such as TCD-TIMIT~\cite{TCDTIMIT}),  other languages (such as Mandarin~\cite{hou2018audio}) and wide variety of speakers and noises~\cite{NOISEX92,snyder2015musan}. An overview of our proposed AV SE model is shown in Figure~\ref{fig:overview}.

In summary, our paper presents six major contributions:
\begin{enumerate}[label=(\roman*)]
    \item A causal, language, noise and speaker independent AV DNN driven model for SE is proposed.
    The model contextually exploits the audio and visual cues, independent of the SNR, to estimate the spectral mask that is used to selectively suppress and enhance each T-F bin.
    \item A first of its kind AV corpus, consisting of high quality binaural speech recorded in real noisy environments such as cafeteria and restaurant, is collected to evaluate the performance of the proposed model in challenging real noisy settings.
    In the literature, a synthetic mixture of clean speech and noise is generally used to evaluate the AV SE methods.
    However, the synthetic mixtures do not depict the real noisy mixtures as in real mixtures the speech is reverberantly mixed with multiple competing noise background sources.
    \item To the best of our knowledge, our paper is first to propose a speaker, noise and language-independent model that generalises on different languages even after training on a small English vocabulary Grid corpus.
    In the literature, it has been shown that a pretrained SE model trained on a single language do not perform well on new languages~\cite{pascual2018language}.
    \item We perform extensive evaluation of our proposed approach, using real noisy ASPIRE corpus, with state-of-the-art A-only SE approaches (including spectral subtraction, linear minimum mean square error) as well as recent DNN based SE models (including SEGAN) using both objective measures (PESQ, SI-SDR, and ESTOI) and subjective MUSHRA (MUlti Stimulus test with Hidden Reference and Anchor) listening tests.
    \item We also study the behaviour of the trained AV model, in terms of objective metrics, when the visual cues are temporarily or permanently absent for random duration of time due to occlusions.
    \item Finally, we critically analyse and compare the performance of A-only CochleaNet model with the AV counterpart to empirically identify the role visual cues plays in the performance of AV model.
    Specifically, we study the behaviour of the A-only and AV models in silent speech regions as well as we conduct listening tests to gauge the model performances on different phonemes. We hypothesise that the model perform better on visually distinguishable phonemes as compared to visually indistinguishable phonemes.
\end{enumerate}

The rest of the paper is organised as follows: Section \ref{sec:related-work} briefly reviews the related work, section \ref{sec:aspire-corpus} presents the ASPIRE corpus collection setup and the postprocessing involved.
Section \ref{sec:cochleanet} presents, CochleaNet, an AV Mask Estimation model for SE.
Section \ref{sec:experiments-and-results} discuss the experimental setup and results.
Section \ref{sec:conclusion} concludes this work and propose future research directions.

\section{Related work}\label{sec:related-work}

This section briefly reviews the related works in the area of A-only and AV SE.

\subsection{Audio-Visual Speech Enhancement}
Ephrat et al.~\cite{ephrat2018looking} proposed a speaker independent AV DNN for complex ratio mask estimation to separate speech from overlapping speech and background noises.
The model is trained on, AVSPEECH, a new large AV corpus consisting of 1500 hours recording with wide variety of languages, people and face poses.
The main limitation, with the aforementioned study is that the model is trained and evaluated on a fixed SNR\@.
Similarly, Gogate et al.~\cite{gogate2018dnn} presented a speaker independent AV DNN for IBM estimation to separate speech from background noises.
However, the model is trained and evaluated using a limited vocabulary Grid corpus~\cite{Grid} and can help in achieving superior performance.
In addition, Hou et al.~\cite{hou2018audio} proposed a speaker-dependent based SE model, trained and evaluated on a single speaker,  that predicts the enhanced spectrogram from the noisy spectrogram using multimodal deep convolutional network.
On the other hand, Gabbay et al.~\cite{gabbay2018visual} trained a convolutional encoder-decoder architecture to estimate the spectrogram of the enhanced speech from noisy speech spectrogram and cropped mouth regions.
However, the model fails to work when the visuals are occluded.
Adeel et al.~\cite{adeel2018lip,adeel2018contextual} proposed a visual-only and AV SE models by integrating an enhanced visually-derived wiener filter (EVWF) and DNN based lip reading regression model.
The preliminary evaluation demonstrated the effectiveness to deal with spectro-temporal variations in any wide variety of noisy environments.
Owens et al.~\cite{owens2018audio} proposed a self-supervised trained network to categorise whether audio and visual streams are temporally aligned.
The model is then used for feature extraction to condition an on/off screen speaker source separation model.
Afouras et al.~\cite{afouras2018deep} trained a DNN to predict both magnitude and phase of denoised speech spectrograms.
Finally, Zhao et al.~\cite{Zhao_2018_ECCV} presented a model to separate the sound of multiple objects from a video (e.g.\ musical instruments).

\subsection{Audio-only Speech Enhancement}

Hershey et al.~\cite{hershey2016deep} proposed deep clustering that exploits discriminatively trained speech embeddings to cluster and separate the different sources.
For time-domain SE, Rethage et al.~\cite{rethage2018wavenet} proposed a non-causal Wavenet based SE model that operates on raw audios to address the invalid short-time fourier transform (STFT) problem~\cite{griffin1984signal} in spectral mask based models.
Similarly, Pandey et al.~\cite{pandey2018new} and Luo et al.~\cite{luo2019conv} proposed a fully-convolutional time-domain SE model that address the shortcomings of separation in the frequency domain, including the decoupling of phase and magnitude, and high latency of calculating the STFT.

A fundamental problem with A-only SE and separation is the label permutation problem~\cite{hershey2016deep} i.e.\ there is no easy way to associate a mixture of audio sources with the corresponding speakers or instruments~\cite{yu2017permutation}. In addition, the main limitation with most of the aforementioned A-only and AV SE approaches is that the developed model are either evaluated on high SNRs ($SNR > 0$  dB) or on a fixed SNR. In addition, none of the aforementioned AV approaches have used an AV speech corpus recorded in real noisy settings for evaluation.

\section{ASPIRE Corpus}\label{sec:aspire-corpus}

In the literature, extensive research has been carried out to develop A-only real noisy mixtures that often consists of speech signal that is reverberantly mixed with multiple competing noise background sources~\cite{CHiME3}.
However, to the best of our knowledge, no such AV corpus recorded in real noisy settings is available.
In this section, we present ASPIRE, a first of its kind, AV speech corpus recorded in real noisy environments (such as cafeteria and restaurant) to support reliable evaluation of AV SE technologies.

\begin{table}[!t]
  \caption{Grid Corpus Sentence Structure e.g. place blue in A 9 soon}
  \label{tab:Grid}
  \centering
  \begin{tabular}{cccccc}
\hline    \multicolumn{1}{c}{\textbf{command}} &
  \multicolumn{1}{c}{\textbf{colour}} &
  \multicolumn{1}{c}{\textbf{preposition}} &
  \multicolumn{1}{c}{\textbf{letter}} &
  \multicolumn{1}{c}{\textbf{digit}} &
  \multicolumn{1}{c}{\textbf{adverb}}\\
    \hline
bin & blue & at & A-Z & 1--9 & again    \\
lay & green & by & minus W & zero & now    \\
place & red & in &  & & please    \\
set & white & with &  & & soon    \\

    \hline
  \end{tabular}

\end{table}

\begin{figure}[!t]
  \centering
  \includegraphics[width=0.72\linewidth]{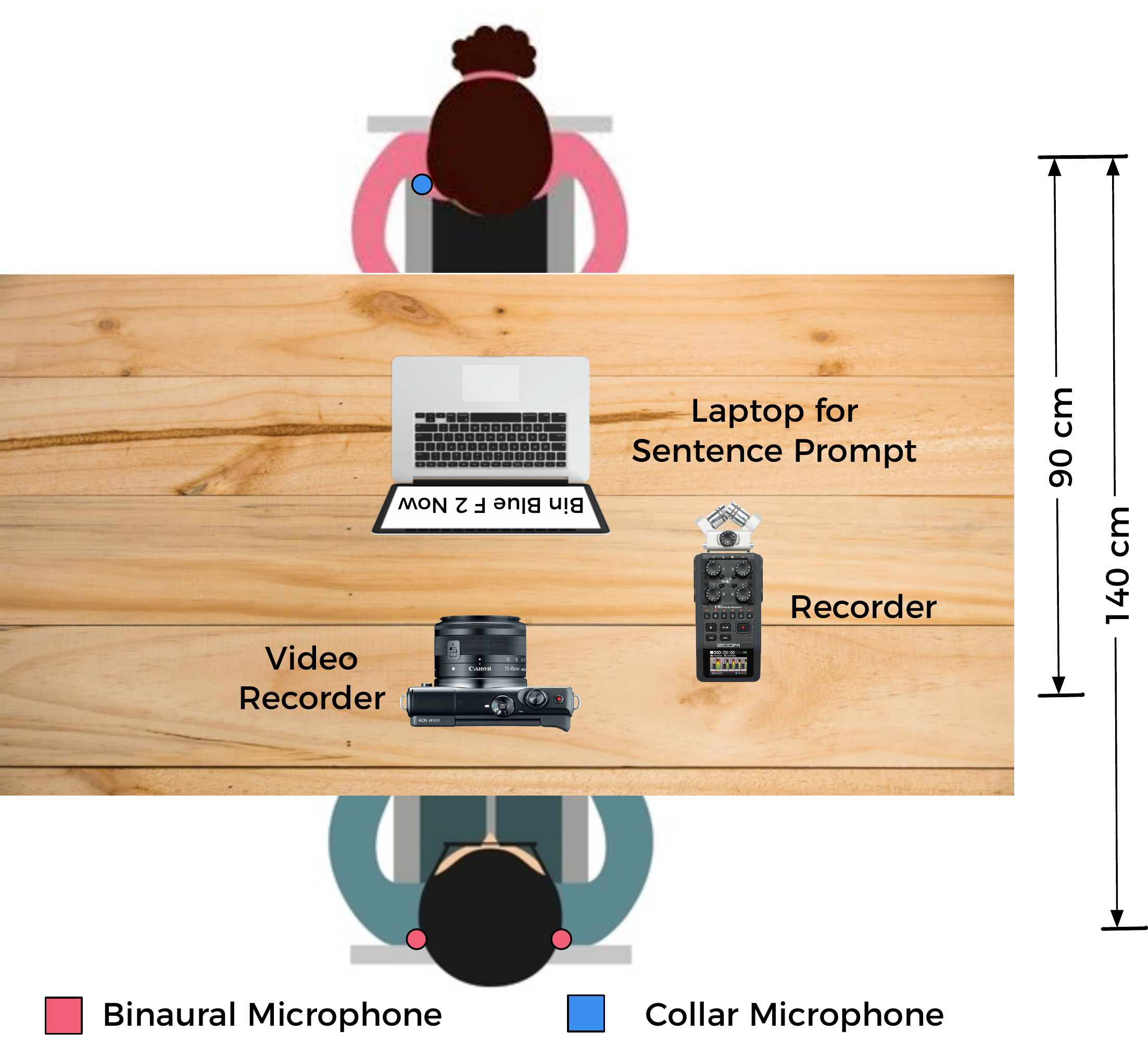}
  \caption{Plan of ASPIRE recording setting showing location of listener, speaker, audio recorder, video recorder, sentence prompter and binaural/collar microphone}
  \label{fig:setupASPIRE}
\end{figure}

\subsection{Sentence design}\label{sec:sentenceDesign}
ASPIRE corpus follows the same sentence format as the AV Grid corpus as shown in Table~\ref{tab:Grid}.
The six words sentence consists of command, colour, preposition, letter, digit and adverb.
The letter "w" was excluded because it is the only multi-syllabic letter.
Each speaker produced all combinations of colour, letter and digit leading to 1000 utterances per talker in both real noisy settings and acoustically isolated booth.
Thus, each talker recorded 2000 utterances.
\subsection{Speaker population}\label{subsec:speaker-population}
Three speakers (one male and two female) contributed to the corpus.
The speakers age ranged from 23 to 55.
All the speakers have spent most of their lives in the United Kingdom and together encompassed a range of mixed English accents.
All the participants were paid for their contribution.
The corpus consists of total 6000 utterances (3000 recorded in real noisy settings, 3000 in acoustically isolated booth).

\subsection{Collection}\label{subsec:collection}
The ASPIRE corpus is recorded in real noisy settings specifically the university cafeteria and restaurant during busy lunch times (11.30 to 1.30) as well as in an acoustically isolated booth.
The recording setup is shown in Figure~\ref{fig:setupASPIRE}.
Apple iPad mini 2, placed at an eye level to avoid noise and distraction from the video apparatus, was used to record the video (the distance between iPad and speaker was 90 centimetres) at 30 frames per second (fps) and 1080p resolution.
A collar microphone was also connected to the iPad.
The high quality binaural audio from speaker is recorded using Zoom H4n pro recorder at a sampling rate of 44100 Hz and binaural microphone.
The listener was wearing the binaural microphone at an approximate distance of 140 centimetres.

The listener and speaker were sitting opposite to each other on the fixed chairs.
Speaker was initially trained with few utterances and the purpose of research is also explained in detail.
Periodic breaks were given to the speakers during the recording to avoid fatigue and each sentence was mandatory to be read correctly without any interruption.
The sentences as detailed in section~\ref{sec:sentenceDesign} were presented to the speaker on a laptop in random order and speaker was allowed to repeat the sentence if the sentence recording is interrupted or sentence is incorrectly uttered.
In addition, the speaker repeated the utterance if any mistake is spotted by the listener.
In all, 2000 utterances per speaker (1000 utterances in real noisy settings and  1000 utterances in the booth) around 2\% and 4\% of the utterances were re-recorded in booth and real noisy settings respectively.

\begin{figure}[!t]
  \centering
  \includegraphics[width=\linewidth]{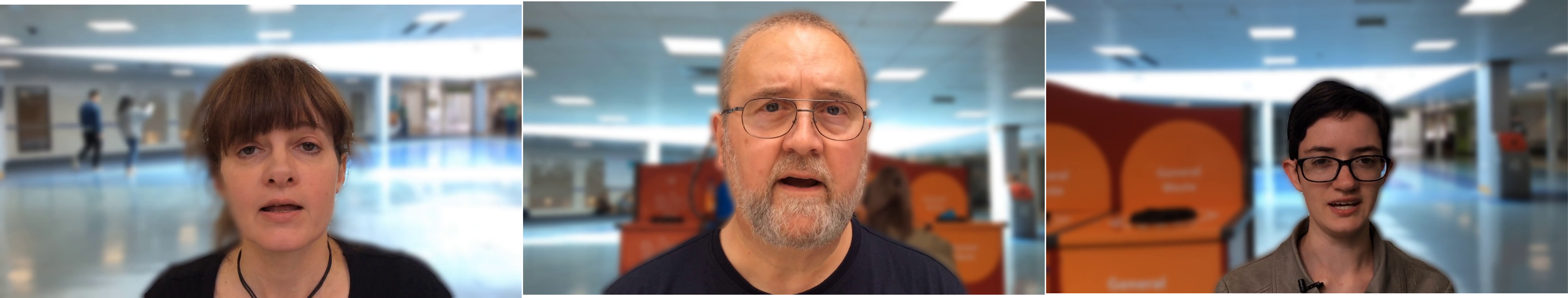}
  \caption{Sample video frames from ASPIRE corpus}
  \label{fig:screenshotsASPIRE}
\end{figure}

\subsection{Postprocessing}\label{subsec:postprocessing}

\paragraph{Audio postprocessing}
Audio and video data were continuously collected throughout a session.
The drift between audio and video data was calculated by synchronising the claps.
The utterance start and end times were identified using Gentle (a robust forced-aligner built on Kaldi), speech recorded from the collar microphone and the presented transcriptions.
Finally, all the segmented utterances were manually checked to correct any additional alignment errors.

\paragraph{Video postprocessing}
The raw videos recorded in busy restaurant and cafeteria consists of a few clearly identifiable people except the speaker itself.
Therefore, to ensure the privacy, we estimate the speaker area for the first frame using a segmentation model and pixelate the non-speaker area for the complete utterance using the estimated segmentation mask.
This is possible because the speaker is sitting in a single position throughout an utterance. Figure~\ref{fig:screenshotsASPIRE} shows some sample video frames from the ASPIRE corpus.

\begin{figure}[!t]
  \centering
  \includegraphics[width=\linewidth]{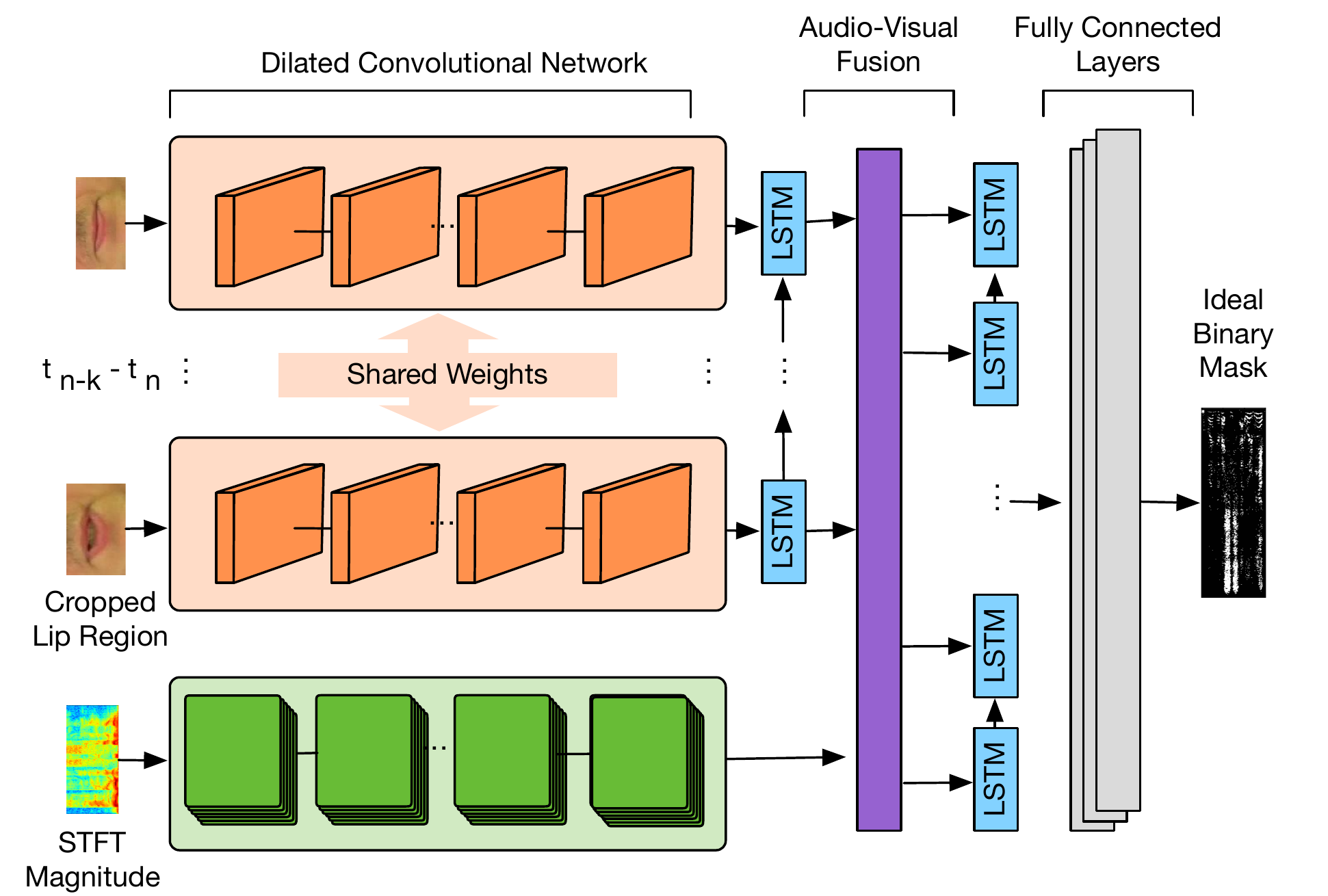}
  \caption{CochleaNet DNN Architecture Overview: Audio-Visual Speech Enhancement}
  \label{fig:cochleanet}
\end{figure}

\section{CochleaNet}\label{sec:cochleanet}
\subsection{Data Representation}\label{subsec:video-and-audio-representation}
\paragraph{Input features} Our model ingests both audio and visual as input.
For batch training, 3 second video clips are considered.
A cropped 80 x 40 lip region is extracted from the video and is used as a visual input (75 cropped lip images assuming 3 second clip recorded at 25 fps).
For audio input, we compute STFT of audio segments and a magnitude spectrogram is used.
The trained model can be applied to both streaming data as well as data of arbitrary lengths during inference time.
\paragraph{Output} The output of our network is an IBM, a multiplicative spectrogram mask, that describes the T-F relationship between clean audio and background noise.
In the literature, it has been shown that the multiplicative masks perform better than direct prediction of time-domain waveform and clean spectrogram magnitudes~\cite{wang2014training,wang2018supervised}.

\begin{table}[!t]

\centering
\caption{Audio Feature Extraction}
\begin{tabular}{lccccc}
\hline

 & conv1 & conv2 & conv3 & conv4 & conv5 \\ \hline
Num filters & 96 & 96 & 96 & 96 & 96 \\
Filter size & 5 x 5 & 5 x 5 & 5 x 5 & 5 x 5 & 1 x 1 \\
Dilation & 1 x 1 & 2 x 1 & 4 x 1 & 8 x 1 & 1 x 1 \\ \hline
\label{tab:audFeatEx}

\end{tabular}
\end{table}

\begin{table}[!t]
\centering
\caption{Visual Feature Extraction}
\begin{tabular}{lccccccc}
\hline
 & conv1 & conv2 & maxpool1 & conv3 & conv4 & maxpool2 & lstm1 \\
 \hline
Num filters & 32 & 48 &  & 64 & 96 &   &  \\
Size & 3 x 3 & 3 x 3 & 2 x 3 & 3 x 3 & 3 x 3 & 2 x 3 & 256 \\
Dilation & 1 x 1 & 1 x 1 &  & 2 x 2 & 3 x 3 &  &  \\
\hline
\label{tab:visFeatEx}

\end{tabular}
\end{table}

\subsection{Network Architecture}\label{subsec:network-architecture}
This section describes the network architecture of the proposed AV SE model.
Figure~\ref{fig:cochleanet} depicts a high-level overview of the multi-stream modules present in the network.
The subsequent subsections describes each module in detail.

\subsubsection{Audio Feature Extraction}
The audio feature extraction consist of dilated convolutional layers as detailed in Table~\ref{tab:audFeatEx}.
Each layer is followed by a ReLU activation for non-linearity.

\subsubsection{Visual Feature Extraction}
The visual feature extraction consist of dilated convolutional, max pooling and long short-term memory (LSTM) layer as detailed in Table~\ref{tab:visFeatEx}.
Each convolutional layer is followed by a ReLU activation for non-linearity.

\subsubsection{Multimodal Fusion}
The visual features are sampled at 25 fps while the audio feature sampling rate is 75 vectors per second (VPS).
Visual features were upsampled to match the audio vector per second rate and to compensate for the sampling rate discrepancies.
This is done using simple repetition of each element 3 times in the temporal dimension.
After upsampling, the audio and visual features are concatenated and fed to a LSTM layer consisting of 622 units.
The LSTM output is then fed to two fully connected layers with 622 neurons and ReLU activation.
The weights of the fully connected layers are shared across the time dimension.
Finally, the extracted features were fed to a fully connected layer with 622 neurons and sigmoid activation.
The binary cross-entropy between the estimated and the actual IBM is used as a loss function.
It is to be noted that, no thresholding was applied to the predicted mask and the sigmoidal outputs were considered as the estimated mask.

\subsection{Speech Resynthesis}\label{subsec:resynthesis}
The model estimates a T-F IBM when a noisy spectrogram and cropped lip images are fed.
The estimated multiplicative spectral mask is applied to the noisy magnitude spectrum.
The masked magnitude is then combined with the noisy phase to get the enhanced speech using ISTFT. Figure~\ref{fig:overview} depicts an overview of speech resynthesis.

\section{Experiments and Results}\label{sec:experiments-and-results}
We qualitatively and quantitatively evaluated our proposed approach with other state-of-the-art A-only and AV SE in real noisy environments and a range of synthetic AV corpora.

\subsection{Synthetic AV Corpora}\label{subsec:synthetic-av-dataset}
This section present the synthetic AV corpora used for training and testing of CochleaNet.
\subsubsection{Grid + ChiMe 3}\label{subsec:gc3}
In our experiments, benchmark Grid corpus~\cite{Grid} is used for the training and evaluation of the proposed framework.
All 33 speakers with 1000 utterances each are considered.
The sentence format is depicted in Table~\ref{tab:Grid}.
The Grid corpus is randomly mixed with non-stationary noises from 3rd CHiME challenge (CHiME 3)\cite{CHiME3}, consisting of bus, cafeteria, street, and pedestrian noises, for SNRs ranging [-12, 9] dB with a step size of 3 dB. It is to be noted that, the trained model is SNR-independent i.e. the utterances at all SNRs were combined for training, and evaluation.
For training, 21000 utterances from 21 speakers were employed.
The model was validated and tested on 4000 and 8000 utterances from 4 and 8 speakers respectively.

\subsubsection{TCD-TIMIT + MUSAN}
For large vocabulary generalisation analysis, we used benchmark TCD-TIMIT~\cite{TCDTIMIT} corpus.
Specifically, 5488 utterances from 56 speakers are mixed with randomly selected non speech noises from MUSAN noises~\cite{snyder2015musan}.
The MUSAN noises includes technical noises (e.g.\ dialtones, fax machine noises etc.) as well as ambient sounds (e.g.\ thunder, wind, footsteps, animal noises etc.).
It to be noted that, all the 5488 utterances were used as a test set to asses the model performance on large vocabulary, speaker and noise independent settings.

\subsubsection{Hou et al. + NOISEX-92}
For language-independent generalisation testing, a Mandarin dataset~\cite{hou2018audio} based on Taiwan Mandarin Hearing in Noise Test (MHINT) with 320 utterances is mixed with randomly selected noise from NOISEX-92~\cite{NOISEX92} consisting of voice babble, factory radio channel and various military noises including fighter jets, engine room, operations room, tank and machine gun.

\subsection{Data Preprocessing}\label{subsec:data-preprocessing}

\subsubsection{Audio Preprocessing}
The audio signals were resampled at 16 kHz and a mono channel is used for processing.
The resampled audio signal was segmented into N 78 millisecond (ms) frames and 17\% increment rate to produce 75 fps.
A hanning window and STFT is applied to produce 622-bin magnitude spectrogram.

\subsubsection{Video Preprocessing}

The Grid and TCD-TIMIT corpora are recorded at 25 fps.
However, the Mandarin dataset~\cite{hou2018audio}, recorded at 30 fps, is downsampled to 25 fps using ffmpeg~\cite{ffmpeg}.
A dlib face detector~\cite{dlib09} is used to locate the faces in each frame of a video clip (75 face cropped images assuming 3 second clip recorded at 25 fps).
The speakers lip images are extracted out of the 25 fps faces video using a minified dlib~\cite{dlib09} model optimised for extracting the lip landmarks.
A region of aspect ratio 1:2 centred at lip-centre is extracted using the lip landmark points.
The extracted region is resized to 40 pixels x 80 pixels and converted to grey scaled image.
It is to be noted that, the lip sequences are extracted at 25 fps and audio features are extracted at 75 VPS.

\subsection{Experimental Setup}\label{subsec:experimental-setup}
For the AV features fusion and mask estimation, the network is trained using TensorFlow library and NVIDIA Titan Xp GPUs. A subset of speakers from Grid ChiME 3 corpus (as described in section~\ref{subsec:synthetic-av-dataset}) are used for training/validation of the neural network and rest of the speakers are used to test the performance of the trained neural network in speaker independent scenario (25\% testing dataset).
The preprocessed training set of Grid ChiME 3 corpus consists of around 25000 utterances, that are split into 21000 and 4000 utterances for training and validation respectively.
It is to be noted that, there was no overlap between the speakers and the noises present in the train, validation and test set for ensuring the speaker and noise independent criteria.
When a missing visual frame is encountered a vector of zeros is used in lieu of the lip image.
The preprocessed dataset consists of cropped lip images and noisy audio spectrogram as input and IBM as an output.
The network is trained using backpropagation with the Adam optimiser~\cite{adam} till the validation error stop decreasing.

\subsection{Objective testing on Synthetic mixtures}\label{subsec:objective-testing-on-synthetic-mixtures}
The quality of re-synthesised speech is evaluated using the following objective metrics for estimating speech intelligibility and aforementioned synthetic AV datasets (section~\ref{subsec:synthetic-av-dataset})

\subsubsection{Perceptual Evaluation of Speech quality (PESQ) comparison} 
\begin{figure}[!t]
	\centering

	\subfloat[Grid + ChiME 3]{\includegraphics[trim=0 30 0 0,clip,width=0.9\textwidth]{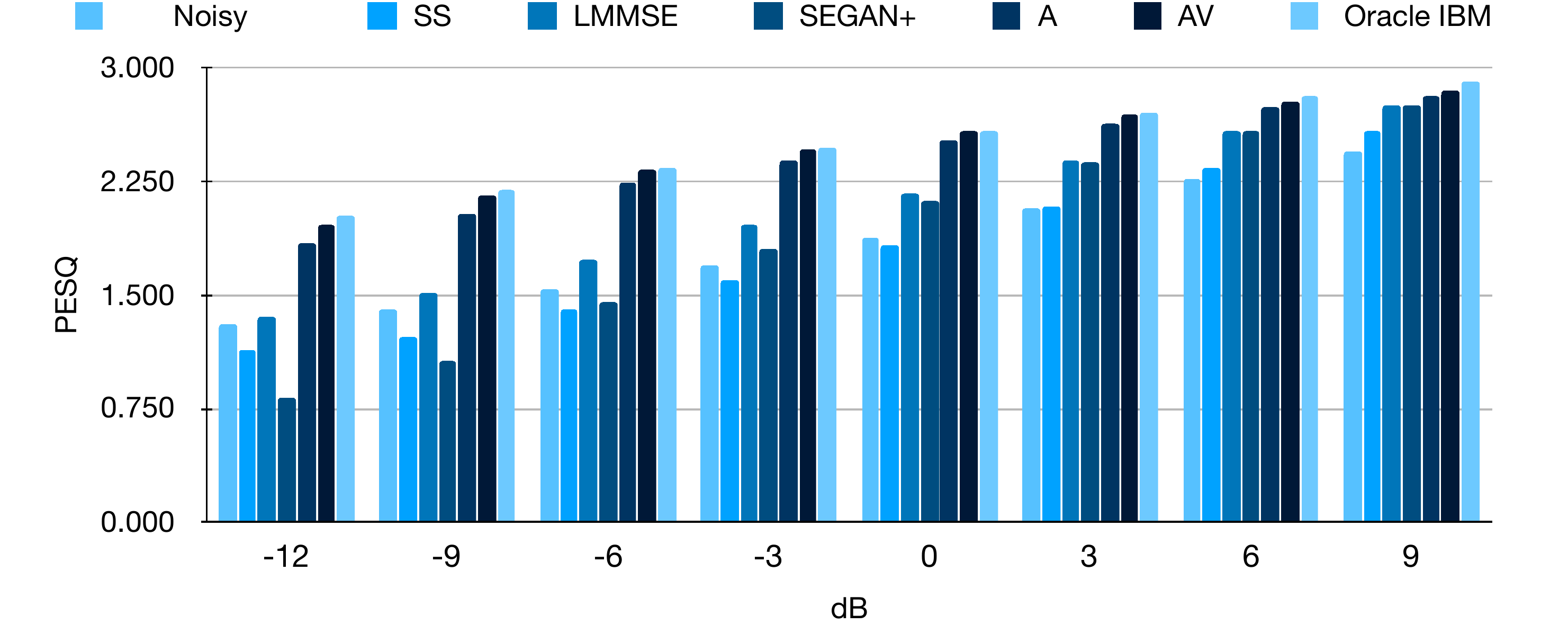}}\\
	\subfloat[TCD + MUSAN]{\includegraphics[trim=0 30 0 30,clip,width=0.9\textwidth]{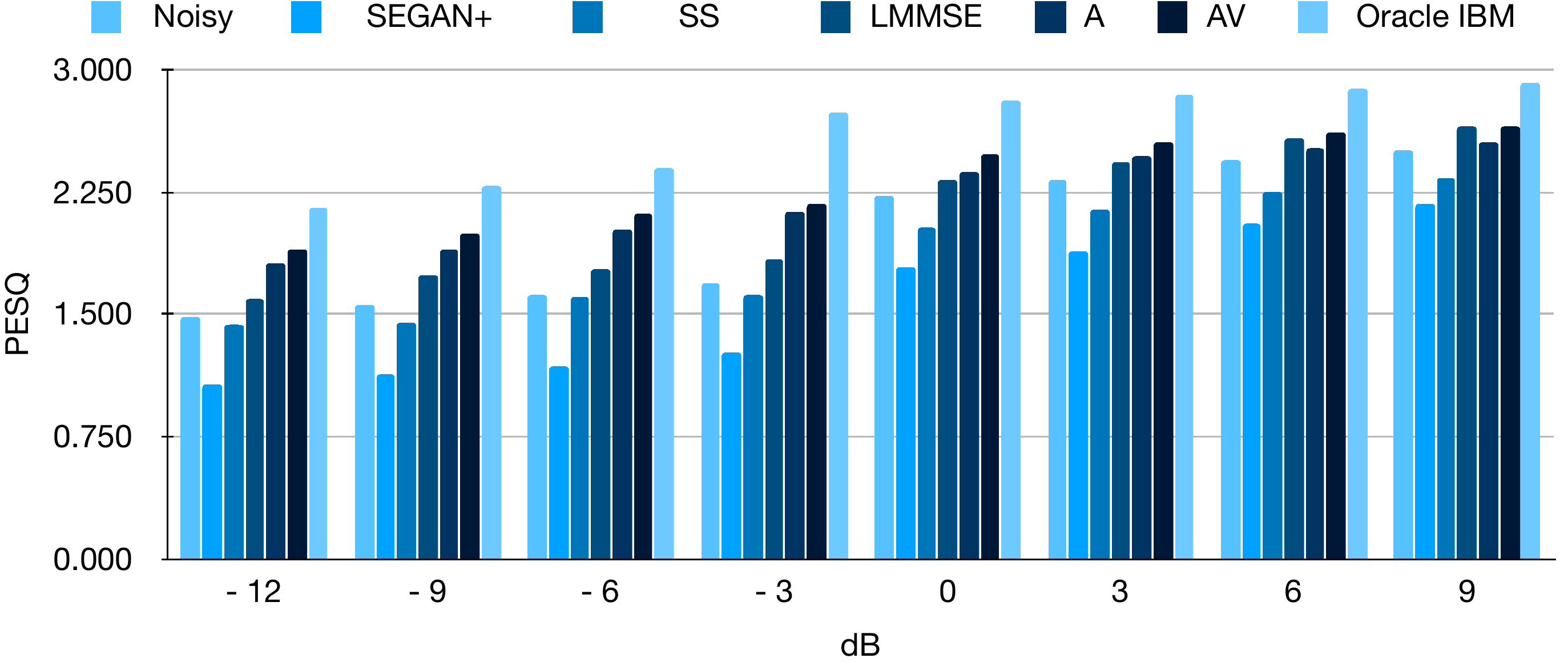}}\\
	\subfloat[Hou et al.
		+ NOISEX-92]{\includegraphics[trim=0 0 0 30,clip,width=0.9\textwidth]{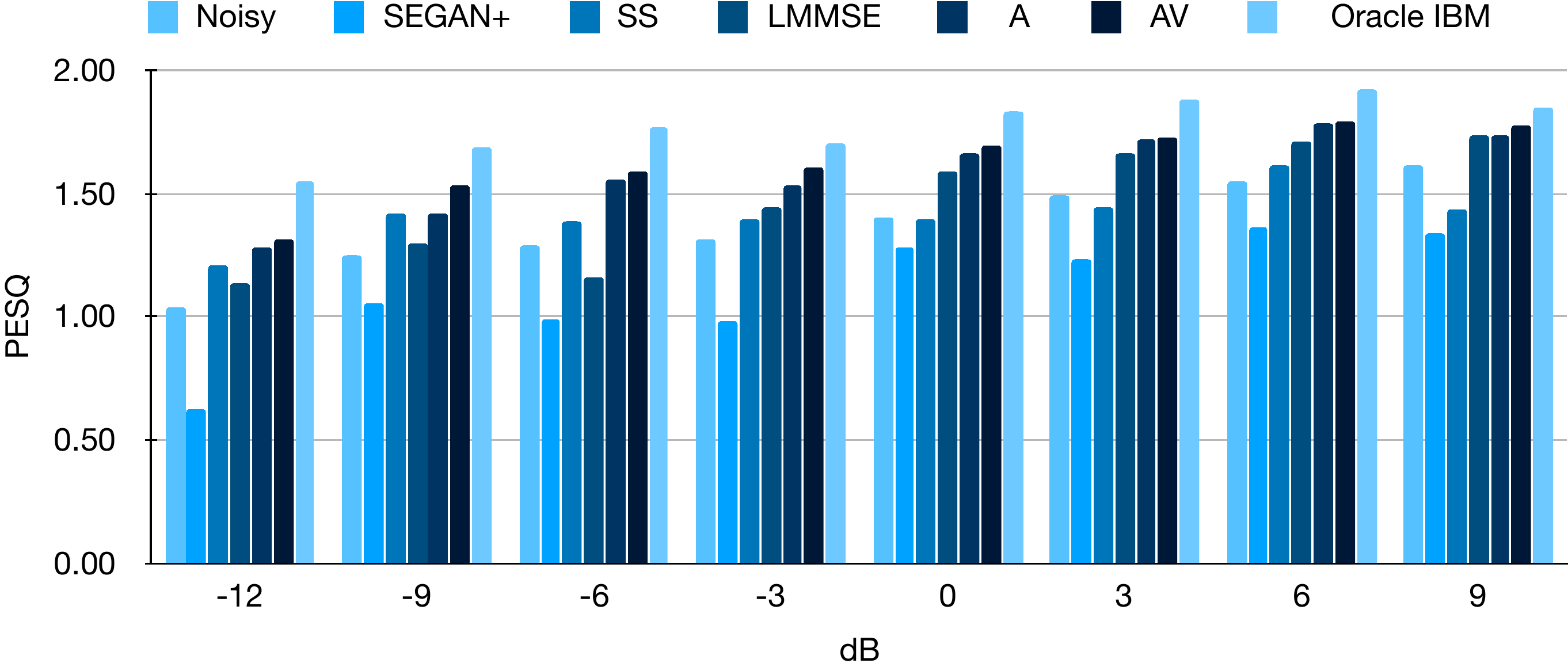}}
	\\
	\caption{PESQ scores for (a) Grid + ChiME3 (b) TCD + MUSAN (c) Hou et al. \cite{hou2018audio} + NOISEx-92 AV dataset computed from the resynthesised speech using SEGAN+~\cite{pascual2017segan}, SS~\cite{boll1979spectral}, LMMSE~\cite{ephraim1985speech}, Audio-only (A) CochleaNet, Audio-Visual (AV) CochleaNet, and Oracle IBM. The reference PESQ for the unprocessed (Noisy) signal is included for relative comparison.}

	\label{fig:pesq_improvement}
\end{figure}

PESQ~\cite{pesq} is one of the most commonly used objective assessment metric in the SE literature and has shown to correlate well with the subjective listening tests~\cite{hu2007evaluation}.
PESQ is computed as a linear combination of the average disturbance value and the average asymmetrical disturbance values between a reference signal and modified signal.
PESQ score ranges from $[-0.50, 4.50]$, indicating the minimum and maximum possible reconstructed speech quality.
The PESQ scores for A-only and AV CochleaNet, SEGAN, SS, and LMMSE with Grid + ChiME 3, TCD TIMIT + MUSAN and Hou et al~\cite{hou2018audio} + NOISEX-92 for different SNRs are presented in Table~\ref{tab:pesqgc3},~\ref{tab:pesqtm},~\ref{tab:pesqac} respectively.
The variety of datasets ensure speaker and noise independent criteria, large vocabulary corpus as well as language-independent scenario.
It is to be noted that, the model trained on Grid + ChiME 3 corpus is used for evaluation.
It can be seen that, at low SNRs, AV CochleaNet and A-only CochleaNet outperformed SS~\cite{boll1979spectral}, LMMSE~\cite{ephraim1985speech}, and SEGAN~\cite{pascual2017segan} based SE methods.
In addition, AV perform better than A-only CochleaNet especially for low SNR ranges (i.e. $SNR < 0$  dB ), where AV CochleaNet model achieved the 1.97, 2.16, and 2.33 PESQ score at SNR levels, of -12dB, -9dB, and -6 dB respectively, as compared to 1.84, 2.04, and 2.24 PESQ score achieved by A-only CochleaNet model for Grid ChiME 3 speaker independent test set.
However, at high SNRs (i.e. $SNR >= 0$  dB) AV slightly outperformed A-only  mask estimation model, where AV CochleaNet achieved 2.58, 2.69, and 2.79 PESQ score at SNR levels, of 0 dB, 3 dB, and 6 dB respectively, as compared to 2.52, 2.63, and 2.73 achieved by A-only CochleaNet model for Grid ChiME 3 speaker independent test set.
The overall PESQ improvement as compared to noisy audio is depicted in Figure~\ref{fig:pesq_improvement}, where AV CochleaNet outperformed the A-only CochleaNet, and achieved near optimal performance (close to an ideal IBM) for Grid ChiME 3 corpus.

\begin{table}[!t]
\centering
\caption{PESQ scores for Grid ChiME 3 speaker independent test set computed from the resynthesised speech using SEGAN+~\cite{pascual2017segan}, SS~\cite{boll1979spectral}, LMMSE~\cite{ephraim1985speech}, Audio-only (A) CochleaNet, Audio-Visual (AV) CochleaNet, and Oracle IBM. The reference PESQ for the unprocessed (Noisy) signal is included for relative comparison.}
\label{tab:pesqgc3}

\begin{tabular}{lcccccccc}
\hline
                & 	-12	    & 	-9	    & 	-6	    & 	-3	    & 	0	    & 	3	    & 	6	    & 	9 \\
 \hline
Noisy           & 	1.30& 	1.40& 	1.54& 	1.70& 	1.87& 	2.07& 	2.27& 	2.45 \\
SEGAN+~\cite{pascual2017segan}      	& 	0.82& 	1.07& 	1.45& 	1.80& 	2.12& 	2.37& 	2.58& 	2.76\\
SS	            & 	1.13& 	1.22& 	1.40& 	1.60& 	1.82& 	2.08& 	2.34& 	2.58\\
LMMSE	        & 	1.36& 	1.51& 	1.73& 	1.96& 	2.17& 	2.39& 	2.58& 	2.75\\
A	            & 	1.84& 	2.04& 	2.24& 	2.39& 	2.52& 	2.63& 	2.73& 	2.81\\
AV	            & 	1.97& 	2.16& 	2.33& 	2.46& 	2.58& 	2.69& 	2.78& 	2.85\\
Oracle IBM	    & 	2.02& 	2.19& 	2.33& 	2.47& 	2.58& 	2.70& 	2.82& 	2.90\\
\hline

\end{tabular}
\end{table}

\begin{table}[!t]
\centering
\caption{PESQ scores for large vocabulary TCD-TIMIT + MUSAN AV dataset computed from the resynthesised speech using SEGAN+~\cite{pascual2017segan}, SS~\cite{boll1979spectral}, LMMSE~\cite{ephraim1985speech}, Audio-only (A) CochleaNet, Audio-Visual (AV) CochleaNet, and Oracle IBM. The reference PESQ for the unprocessed (Noisy) signal is included for relative comparison.}
\label{tab:pesqtm}

\begin{tabular}{lcccccccc}
\hline
               & 	-12	    & 	-9	    & 	-6	    & 	-3	    & 	0	    & 	3	    & 	6	    & 	9 \\
 \hline
Noisy 		            &	1.48	&	1.56	&	1.62	&	1.68	&	2.23	&	2.32	&	2.44	&	2.50\\
SEGAN+~\cite{pascual2017segan}		            &	1.07	&	1.13	&	1.18	&	1.27	&	1.78	&	1.89	&	2.05	&	2.17\\
 SS 		            &	1.43	&	1.44	&	1.60	&	1.62	&	2.03	&	2.14	&	2.25	&	2.33\\
LMMSE		            &	1.60	&	1.73	&	1.77	&	1.83	&	2.32	&	2.43	&	2.57	&	2.65\\
A		                &	1.81	&	1.90	&	2.02	&	2.12	&	2.37	&	2.46	&	2.51	&	2.56\\
AV + No Visuals		    &	1.81	&	1.94	&	2.04	&	2.11	&	2.40	&	2.47	&	2.54	&	2.58\\
AV		                &	1.89	&	1.99	&	2.11	&	2.17	&	2.47	&	2.55	&	2.62	&	2.65\\
Oracle IBM		        &	2.55	&	2.58	&	2.69	&	2.73	&	2.81	&	2.84	&	2.88	&	2.92\\
\hline

\end{tabular}
\end{table}

  \begin{table}[!t]
\centering
\caption{PESQ scores for Hou et al. \cite{hou2018audio} + NOISEX92 AV language-independent dataset computed from the resynthesised speech using SEGAN+~\cite{pascual2017segan}, SS~\cite{boll1979spectral}, LMMSE~\cite{ephraim1985speech}, Audio-only (A) CochleaNet, Audio-Visual (AV) CochleaNet, and Oracle IBM. The reference PESQ for the unprocessed (Noisy) signal is included for relative comparison.}
\label{tab:pesqac}

\begin{tabular}{lcccccccc}
\hline
                & 	-12	    & 	-9	    & 	-6	    & 	-3	    & 	0	    & 	3	    & 	6	    & 	9 \\
 \hline
Noisy	            &	1.04	&	1.25	&	1.29	&	1.31	&	1.40	&	1.49	&	1.71	&	1.64 \\
SEGAN+~\cite{pascual2017segan}	            &	0.63	&	1.06	&	0.99	&	0.98	&	1.28	&	1.23	&	1.36	&	1.34 \\
SS	                &	1.21	&	1.42	&	1.39	&	1.40	&	1.40	&	1.44	&	1.61	&	1.44 \\
LMMSE	            &	1.14	&	1.30	&	1.16	&	1.45	&	1.59	&	1.66	&	1.71	&	1.74 \\
  A 	            &	1.28	&	1.42	&	1.56	&	1.53	&	1.66	&	1.72	&	1.79	&	1.74 \\
AV + No Visuals	    &	1.23	&	1.45	&	1.44	&	1.46	&	1.66	&	1.68	&	1.74	&	1.75 \\
 AV	                &	1.32	&	1.53	&	1.54	&	1.56	&	1.70	&	1.72	&	1.77	&	1.78 \\
  Oracle IBM    	&	1.55	&	1.69	&	1.77	&	1.70	&	1.83	&	1.88	&	1.92	&	1.85 \\
\hline

\end{tabular}
\end{table}

\begin{figure}[!t]
	\centering

	\subfloat[Grid + ChiME 3]{\includegraphics[trim=0 30 0 0,clip,width=0.9\textwidth]{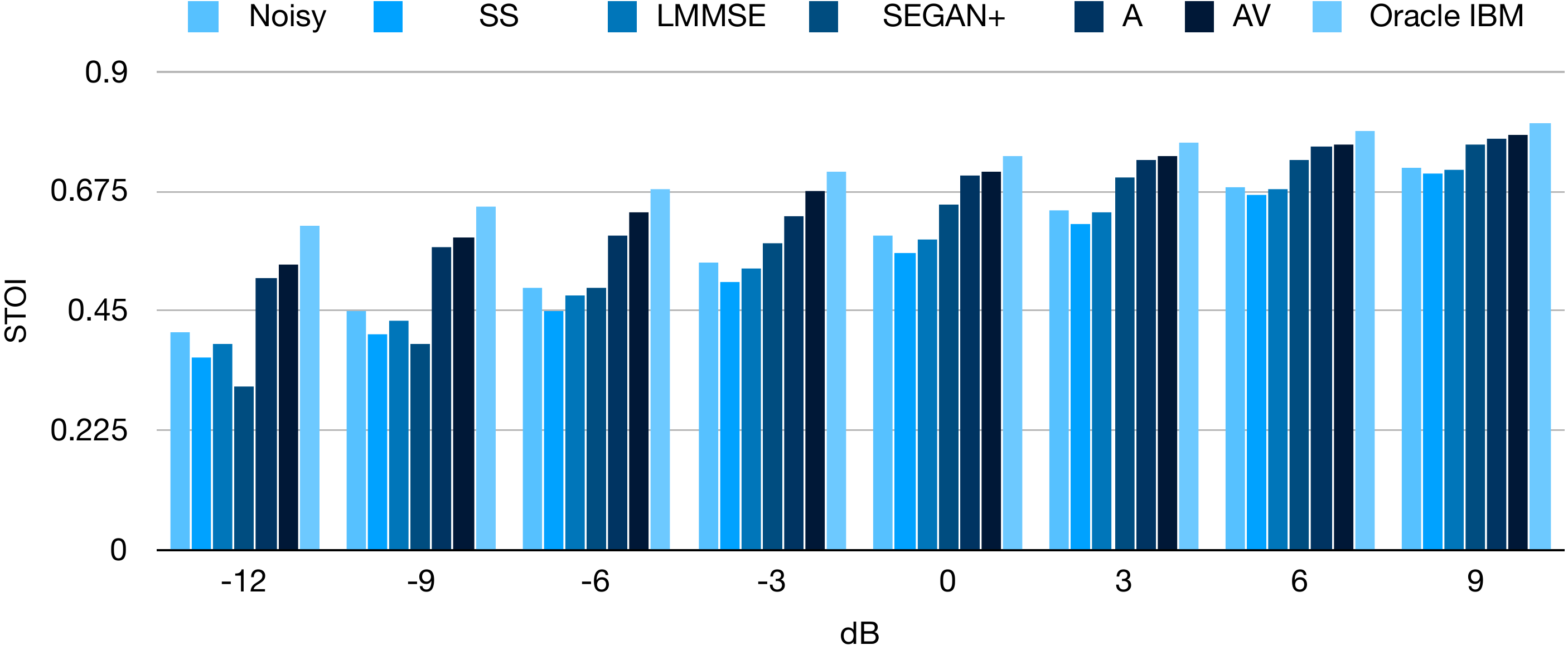}}\\
	\subfloat[TCD + MUSAN]{\includegraphics[trim=0 30 0 30,clip,width=0.9\textwidth]{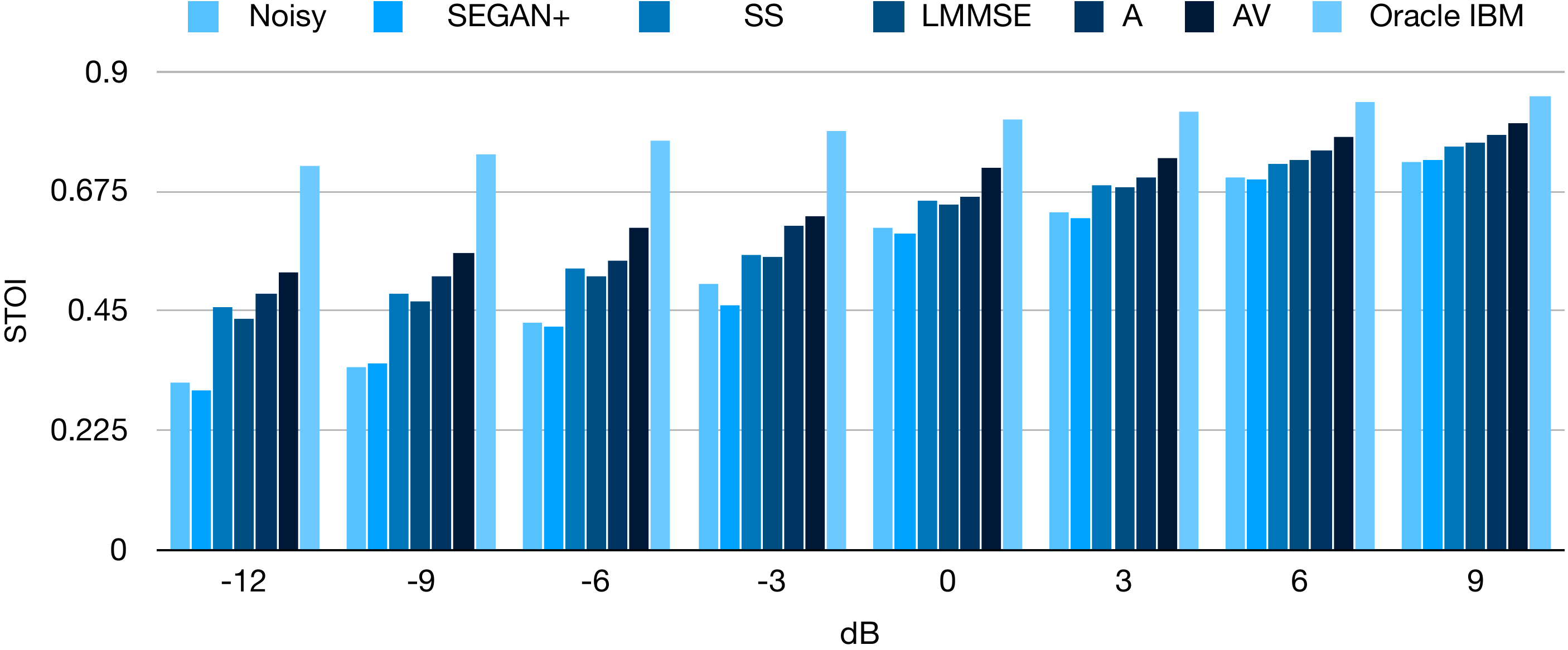}}\\
	\subfloat[Hou et al.
		+ NOISEX-92]{\includegraphics[trim=0 0 0 30,clip,width=0.9\textwidth]{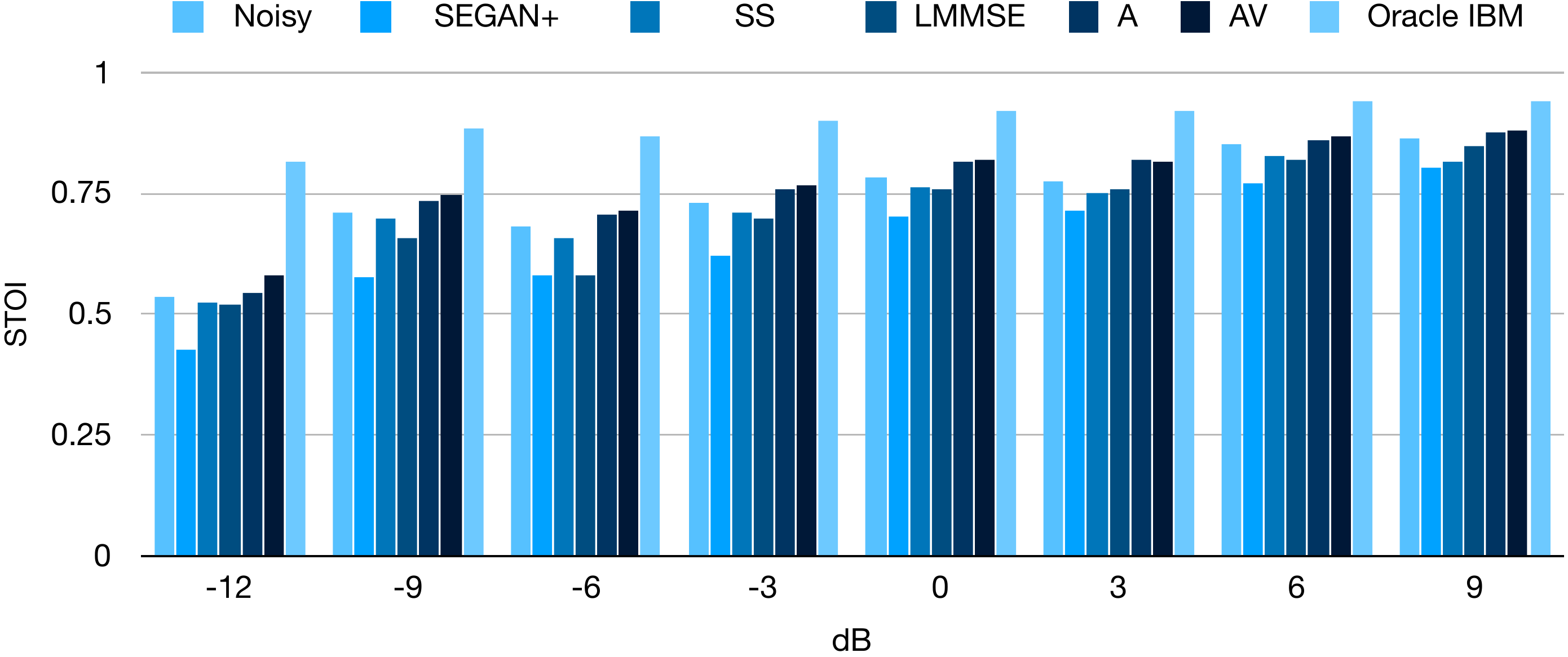}}
	\\
	    \caption{STOI scores for (a) Grid + ChiME3 (b) TCD + MUSAN (c) Hou et al. \cite{hou2018audio} + NOISEX-92 AV dataset computed from the resynthesised speech using SEGAN+~\cite{pascual2017segan}, SS~\cite{boll1979spectral}, LMMSE~\cite{ephraim1985speech}, Audio-only (A) CochleaNet, Audio-Visual (AV) CochleaNet, and Oracle IBM. The reference STOI for the unprocessed (Noisy) signal is included for relative comparison.}

	\label{fig:stoi_improvement}
\end{figure}

\subsubsection{Short Term Objective Intelligibility (STOI) comparison} STOI is another benchmark objective evaluation metric used for speech intelligibility that shows high correlation with subjective listening test scores~\cite{stoi}. The correlation of short-time temporal envelopes between the clean and modified speech is calculated in STOI with values ranging from $[0, 1]$, and a higher value indicates better intelligibility.
The STOI scores for A-only and AV CochleaNet, SEGAN, SS, and LMMSE with Grid + ChiME 3, TCD TIMIT + MUSAN and Hou et al~\cite{hou2018audio} + NOISEX-92 for different SNRs are presented in Fig~\ref{fig:stoi_improvement}.
It can be seen that, at low SNRs, AV CochleaNet and A-only CochleaNet outperformed SS~\cite{boll1979spectral}, LMMSE~\cite{ephraim1985speech}, SEGAN~\cite{pascual2017segan} based SE methods.
In addition, AV perform better than A-only model especially for low SNR ranges (i.e. $SNR < 0$  dB ), where AV CochleaNet model achieved the STOI scores of 0.521, 0.560, and 0.607 at SNR levels, of -12dB, -9dB, and -6 dB respectively, as compared to 0.483,  0.513, and 0.544 achieved by A-only CochleaNet model for  Hou et al~\cite{hou2018audio} + NOISEX-92 language-independent test set.
However, at high SNRs (i.e. $SNR >= 0$  dB) AV slightly outperformed A-only  mask estimation model, where AV CochleaNet achieved STOI scores of 0.719, 0.739, and 0.776 at SNR levels, of 0 dB, 3 dB, and 6 dB respectively, as compared to 0.665, 0.701, and 0.752  achieved by A-only CochleaNet model for Hou et al~\cite{hou2018audio} + NOISEX-92 language-independent test set.

\subsubsection{Scale-Invariant Signal-to-Distortion Ratio (SI-SDR) comparison} 
\begin{figure}[!t]
	\centering

	\subfloat[Grid + ChiME 3]{\includegraphics[trim=0 30 0 0,clip,width=0.9\textwidth]{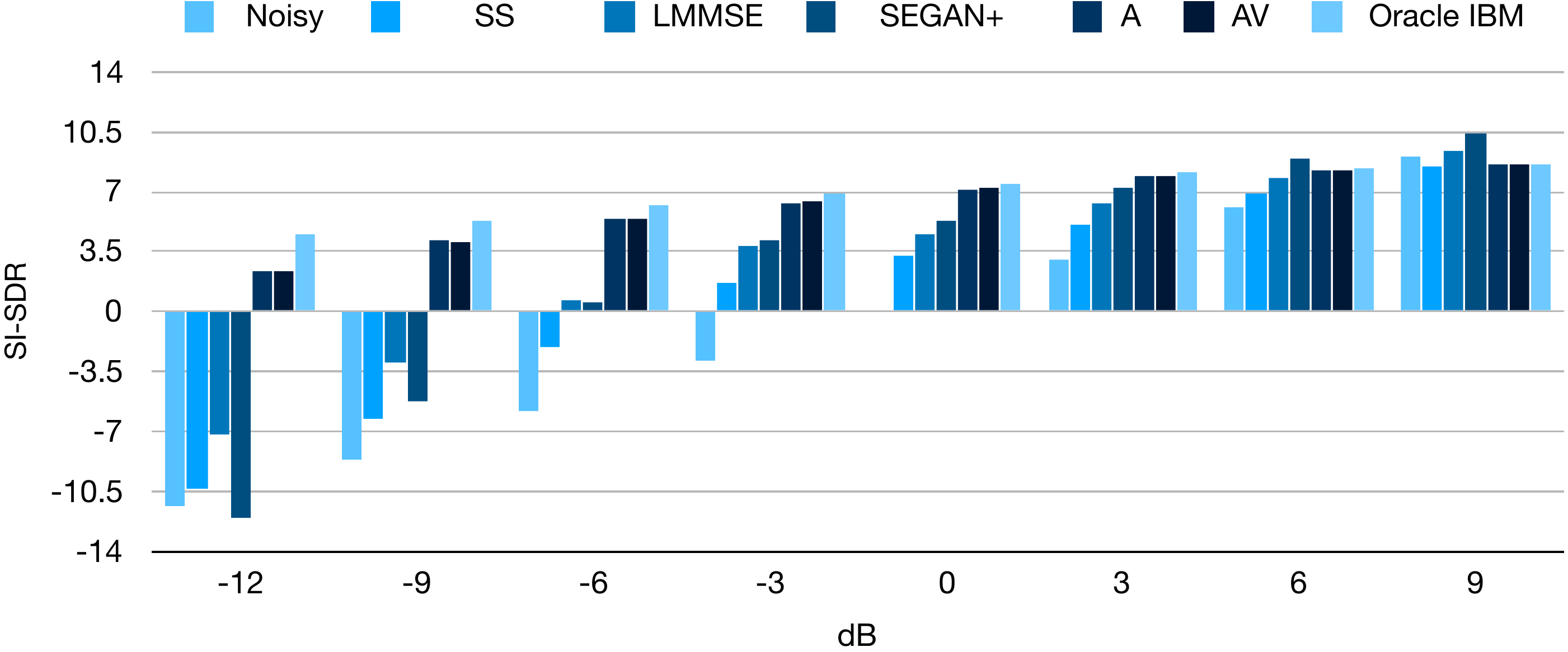}}\\
	\subfloat[TCD + MUSAN]{\includegraphics[trim=0 30 0 30,clip,width=0.9\textwidth]{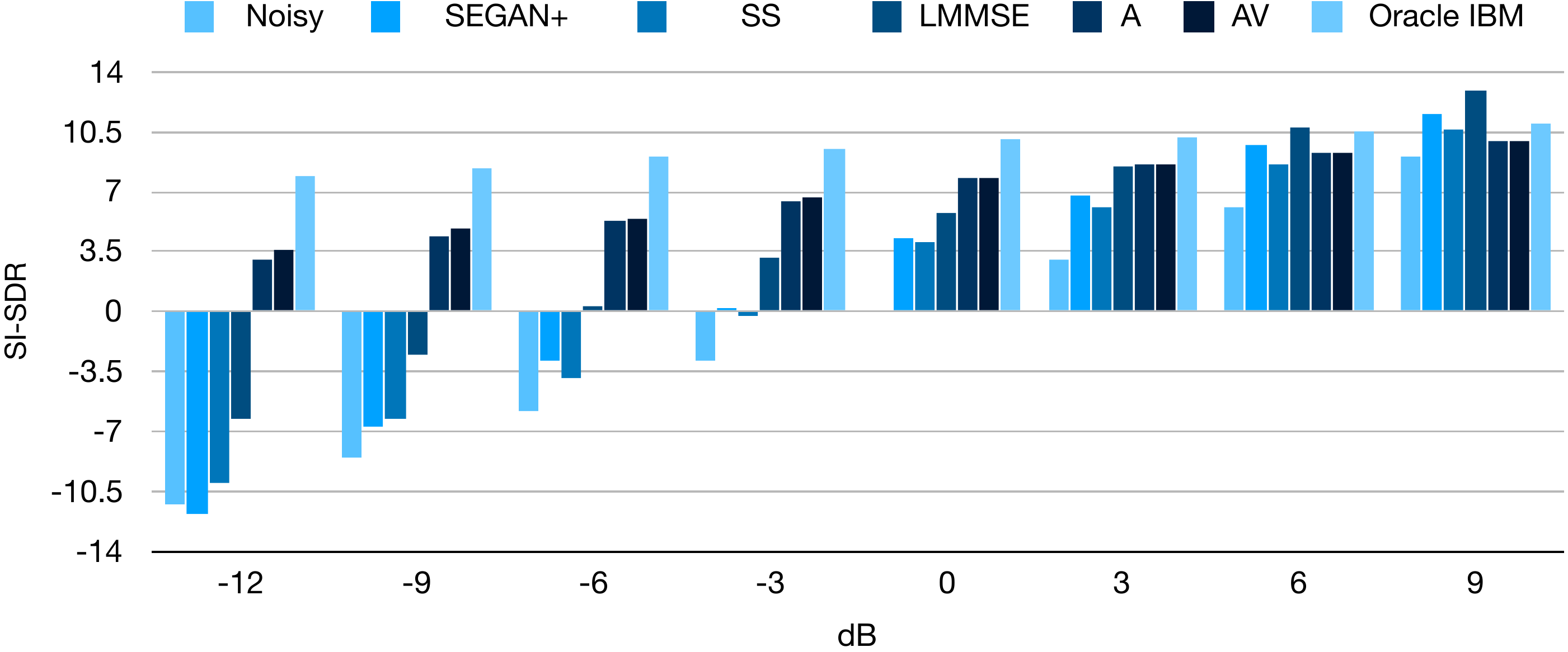}}\\
	\subfloat[Hou et al.
		+ NOISEX-92]{\includegraphics[trim=0 0 0 30,clip,width=0.9\textwidth]{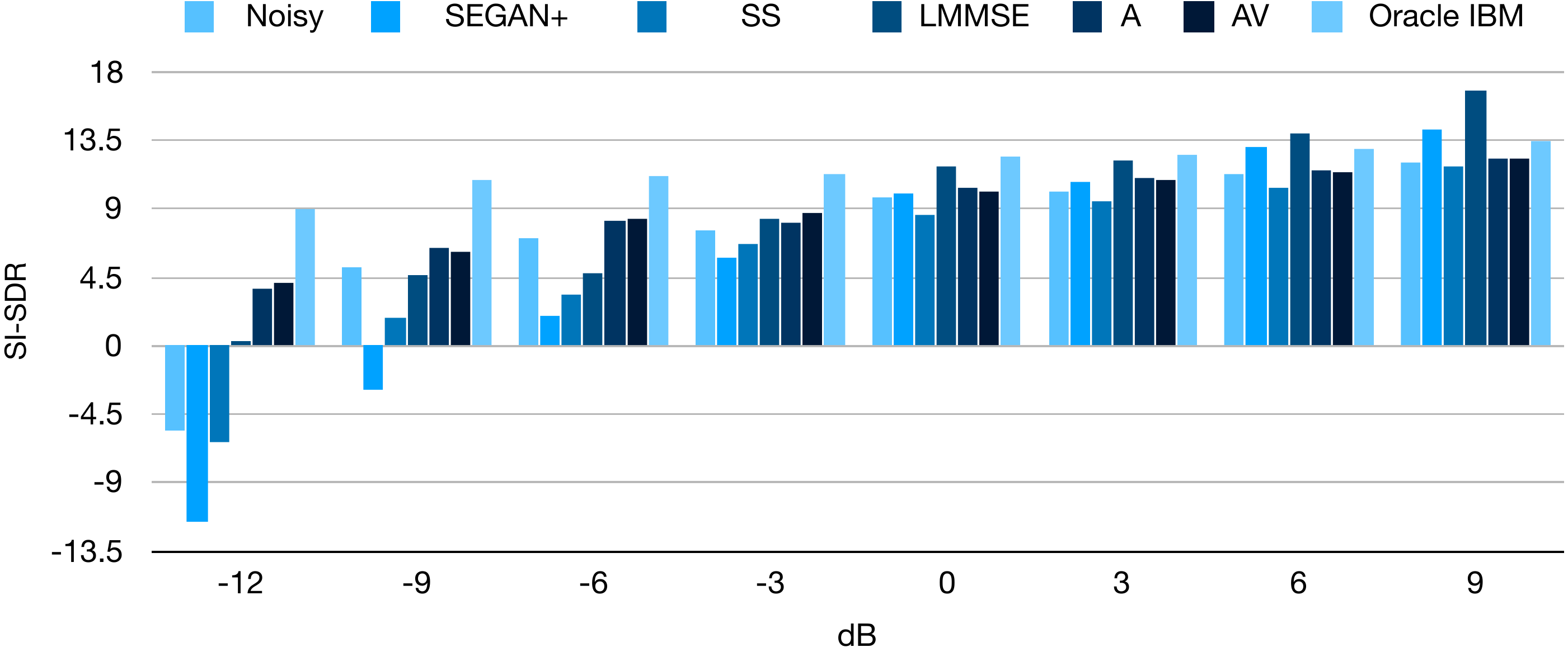}}
	\\
    \caption{SI-SDR scores for (a) Grid + ChiME3 (b) TCD + MUSAN (c) Hou et al. \cite{hou2018audio} + NOISEX-92 AV dataset computed from the resynthesised speech using SEGAN+~\cite{pascual2017segan}, SS~\cite{boll1979spectral}, LMMSE~\cite{ephraim1985speech}, Audio-only (A) CochleaNet, Audio-Visual (AV) CochleaNet, and Oracle IBM. The reference SI-SDR for the unprocessed (Noisy) signal is included for relative comparison.}
	\label{fig:sisdr_improvement}
\end{figure}

SI-SDR~\cite{sisdr} is slightly modified scale invariant version of SDR. SDR is one of the standard speech separation evaluation metric that measures the amount of distortion introduced by the separated signal and is defined as the ratio between clean signal energy and distortion energy.
The higher SDR values indicate better speech separation performance.
The SI-SDR scores for A-only and AV CochleaNet, SEGAN, SS, and LMMSE with Grid + ChiME 3, TCD TIMIT + MUSAN and Hou et al~\cite{hou2018audio} + NOISEX-92 for different SNRs are presented in Fig~\ref{fig:sisdr_improvement} respectively.
It can be seen that, at low SNRs, AV CochleaNet and A-only CochleaNet outperformed SS~\cite{boll1979spectral}, LMMSE~\cite{ephraim1985speech}, SEGAN~\cite{pascual2017segan} based SE methods.
In addition, AV perform better than A-only  mask estimation model especially for low SNR ranges (i.e. $SNR < 0$  dB ), where AV CochleaNet model achieved the SI-SDR scores of 3.62, 4.80, and 5.41  at SNR levels, of -12dB, -9dB, and -6 dB respectively, as compared to 3.04, 4.41, and 5.29 achieved by A-only CochleaNet model for  TCD-TIMIT + MUSAN speaker independent and large vocabulary test set.
However, at high SNRs (i.e. $SNR >= 0$  dB) AV slightly outperformed A-only  mask estimation model, where AV CochleaNet achieved SI-SDR scores of 7.77, 8.64, and 9.31 at SNR levels, of 0 dB, 3 dB, and 6 dB respectively, as compared to 7.76, 8.62, and 9.27 achieved by A-only CochleaNet model for TCD-TIMIT + MUSAN speaker independent and large vocabulary test set.

\begin{figure}[!t]
	\centering
	\subfloat[Noisy]{\includegraphics[width=0.33\textwidth]{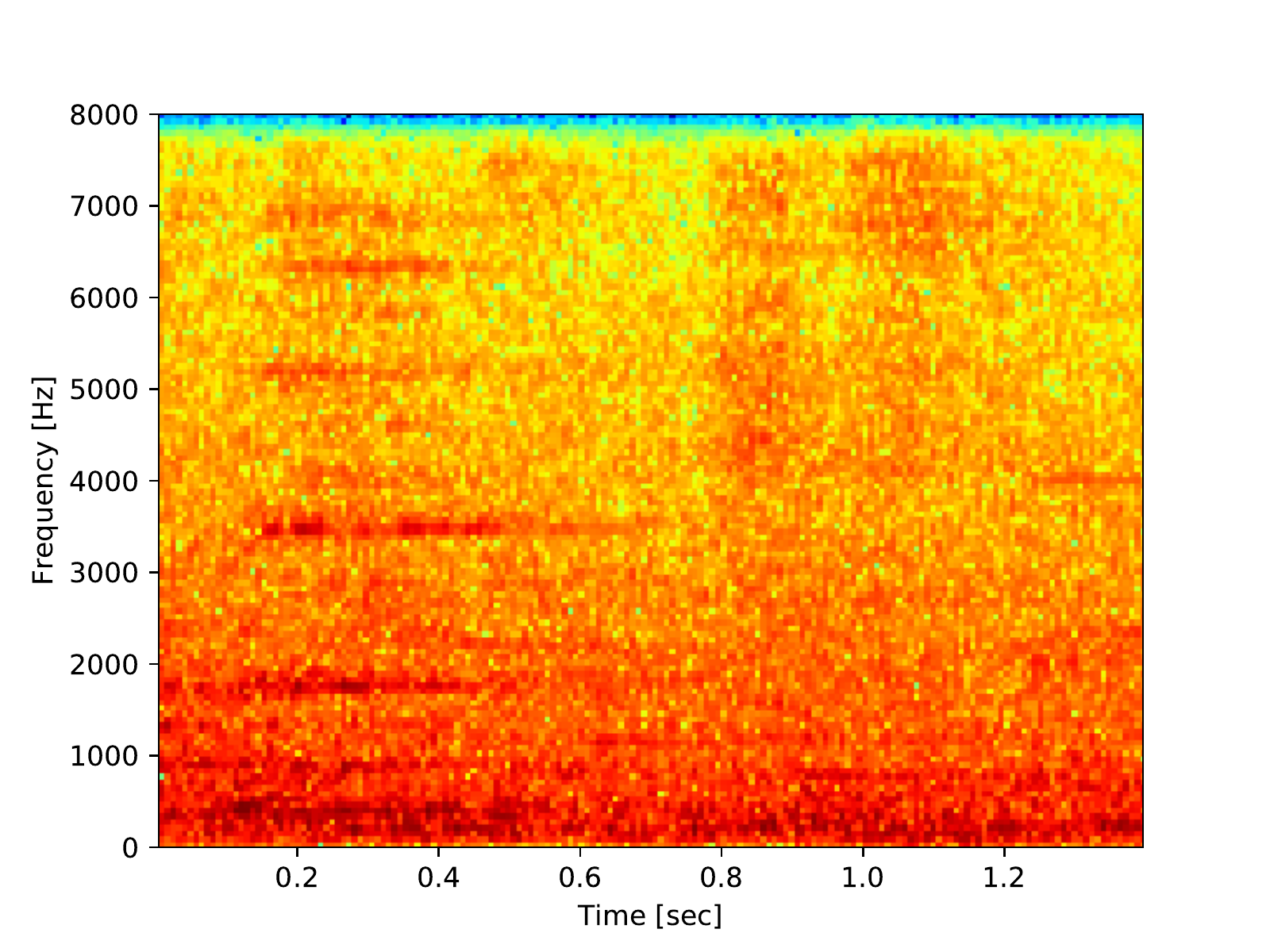}}
	\subfloat[SS]{\includegraphics[width=0.33\textwidth]{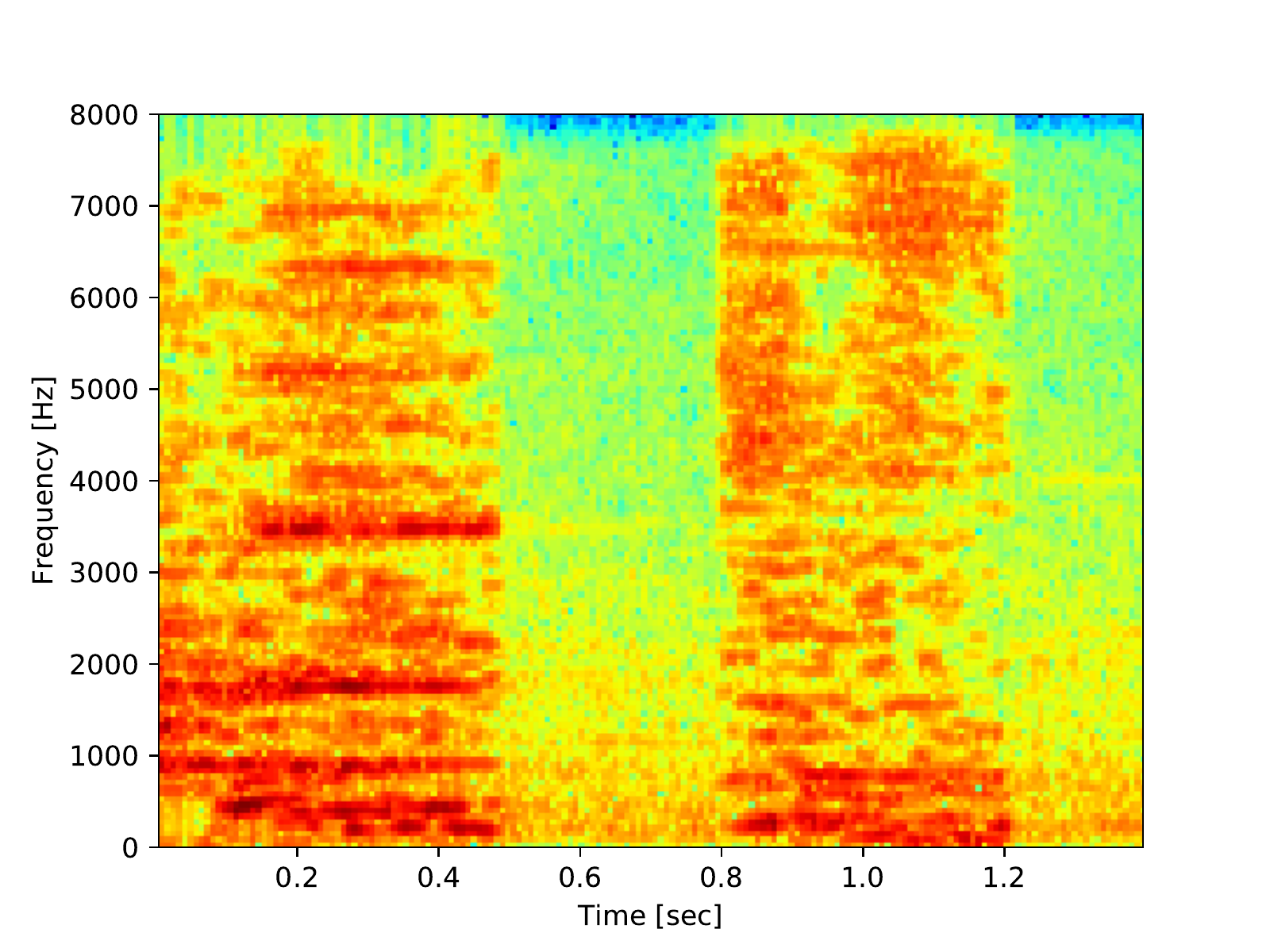}}
	\subfloat[LMMSE]{\includegraphics[width=0.33\textwidth]{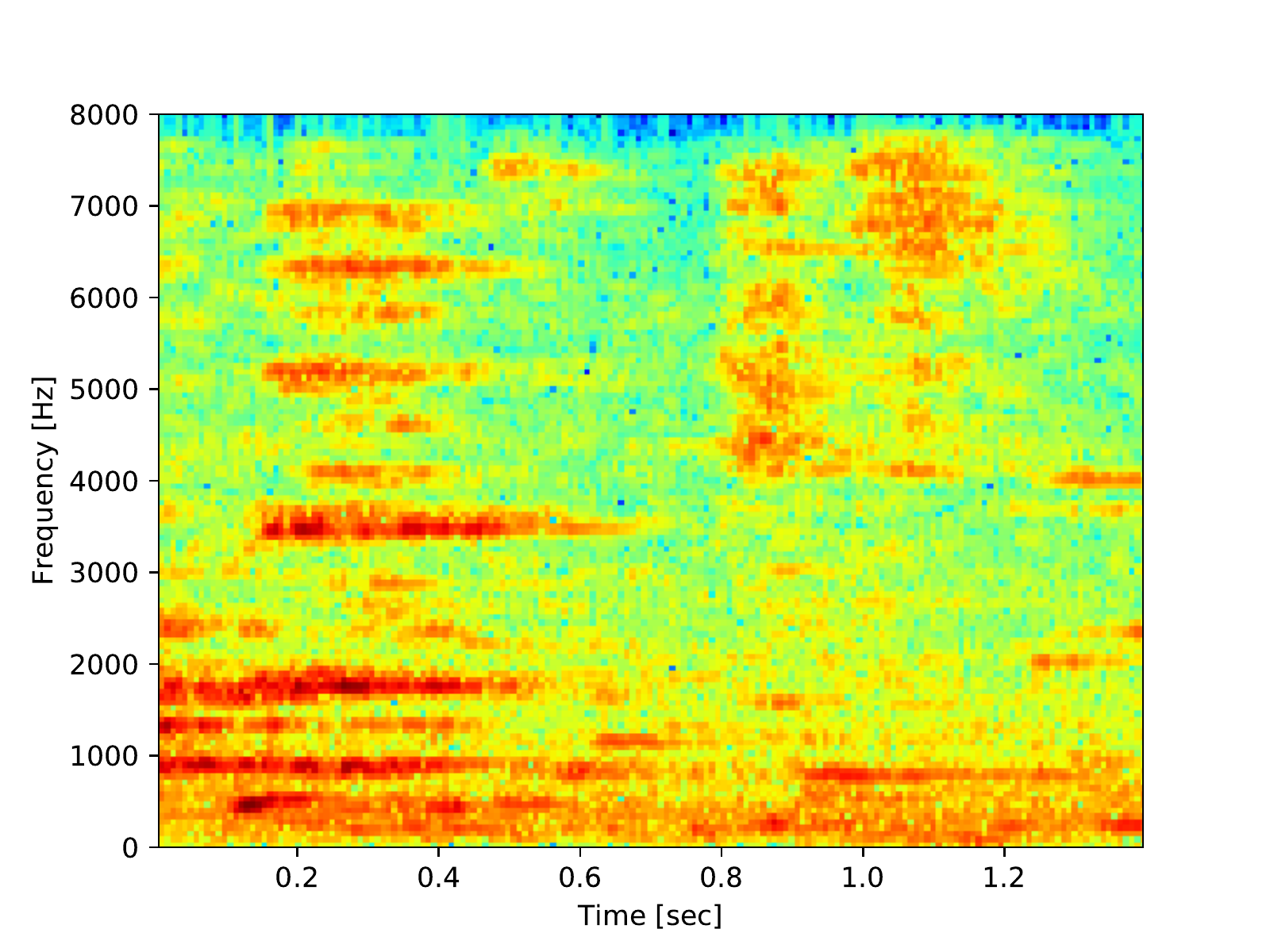}}\\
	\subfloat[SEGAN+~\cite{pascual2017segan}]{\includegraphics[width=0.33\textwidth]{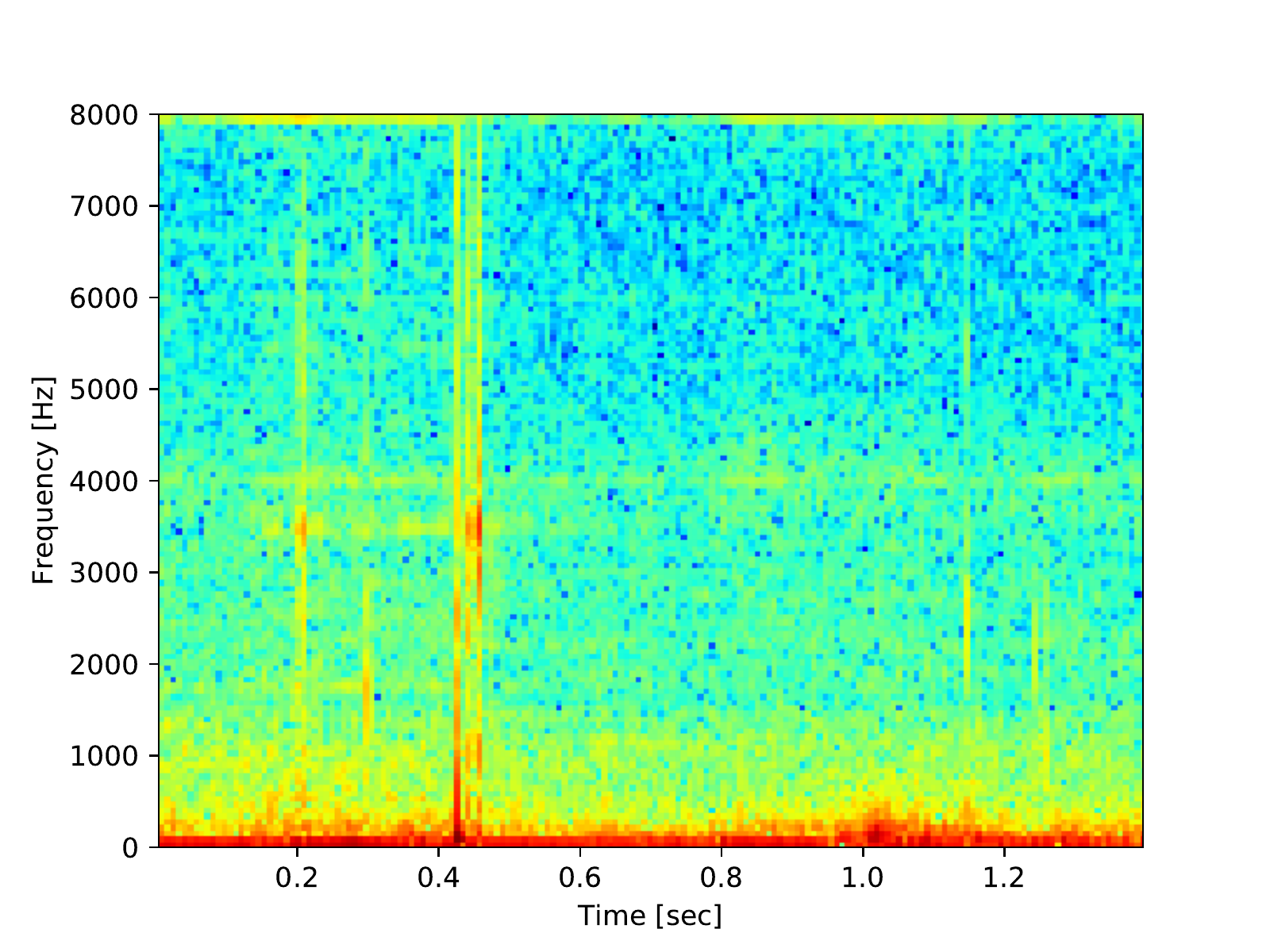}}
	\subfloat[A-Only CochleaNet ]{\includegraphics[width=0.33\textwidth]{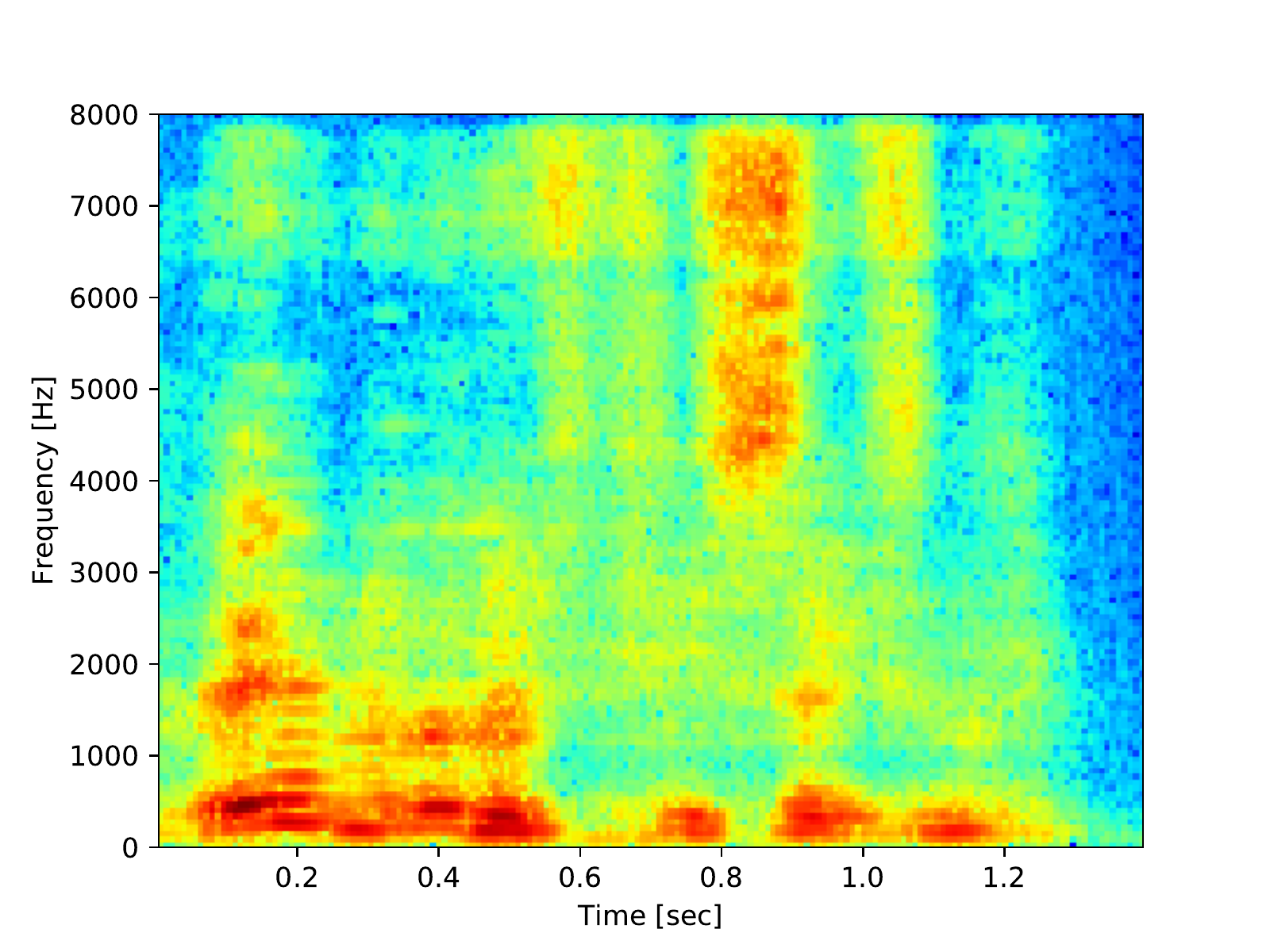}}
	\subfloat[AV CochleaNet]{\includegraphics[width=0.33\textwidth]{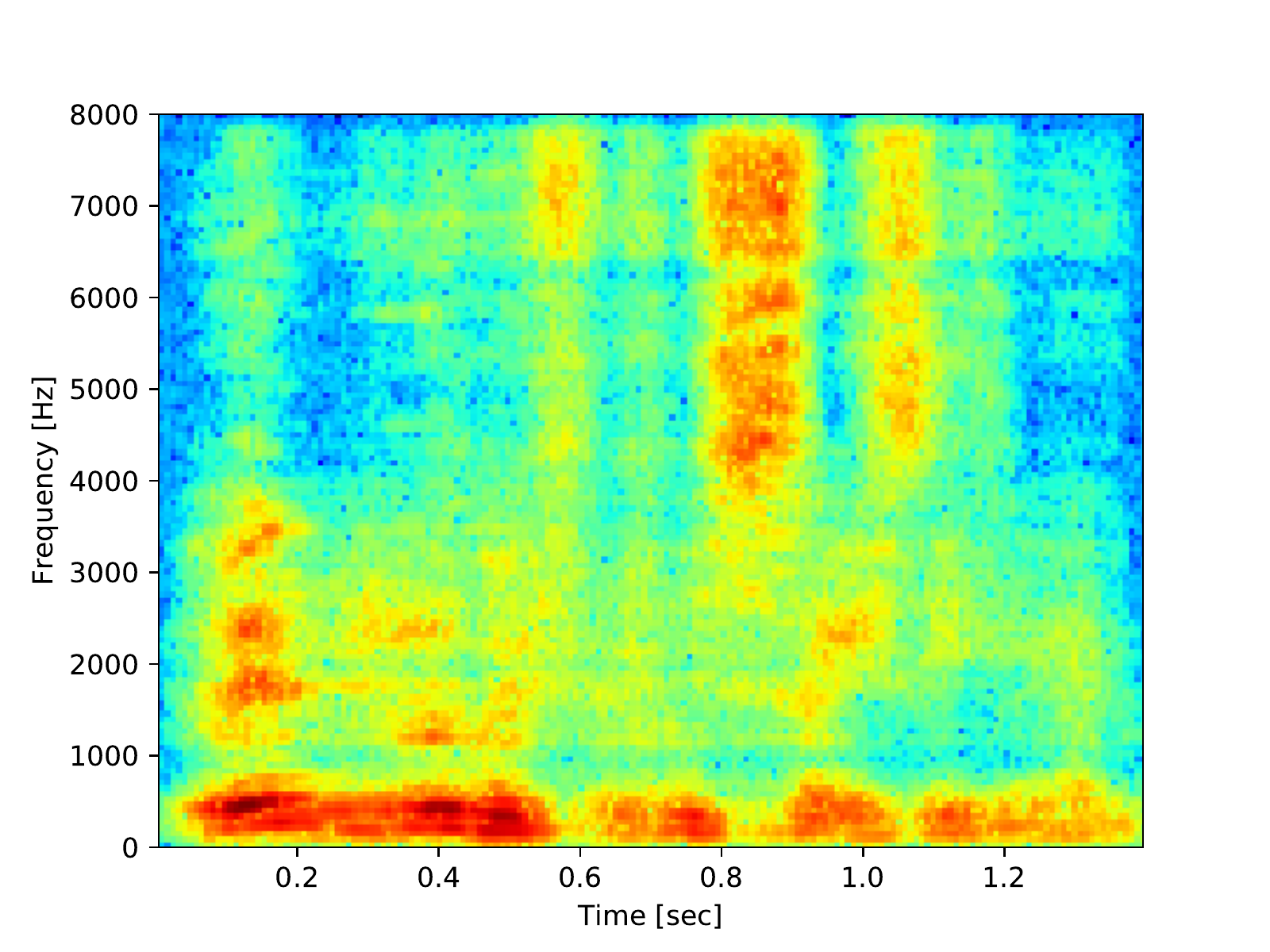}} \\
	\subfloat[Oracle IBM]{\includegraphics[width=0.33\textwidth]{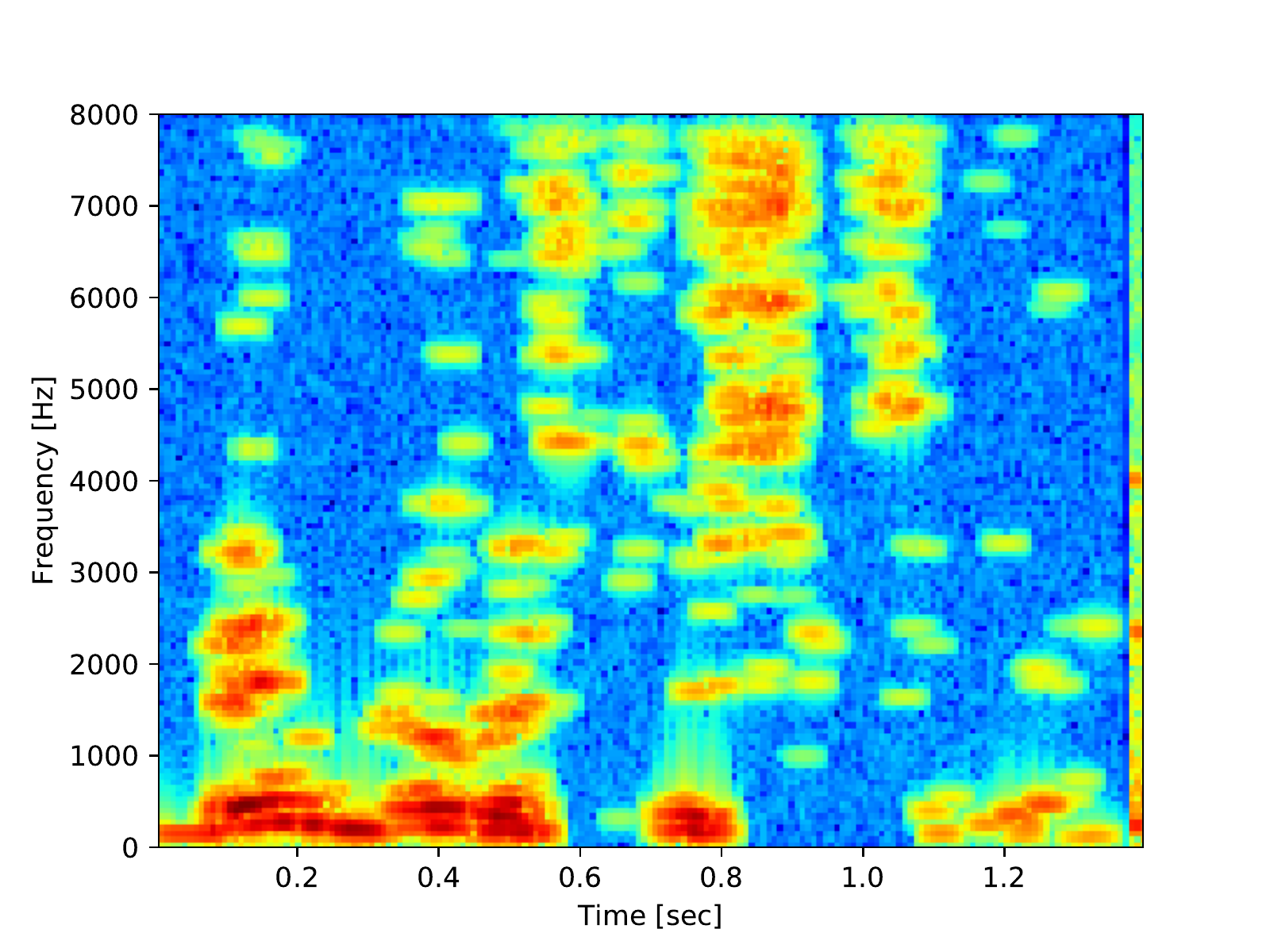}}
	\subfloat[Clean]{\includegraphics[width=0.33\textwidth]{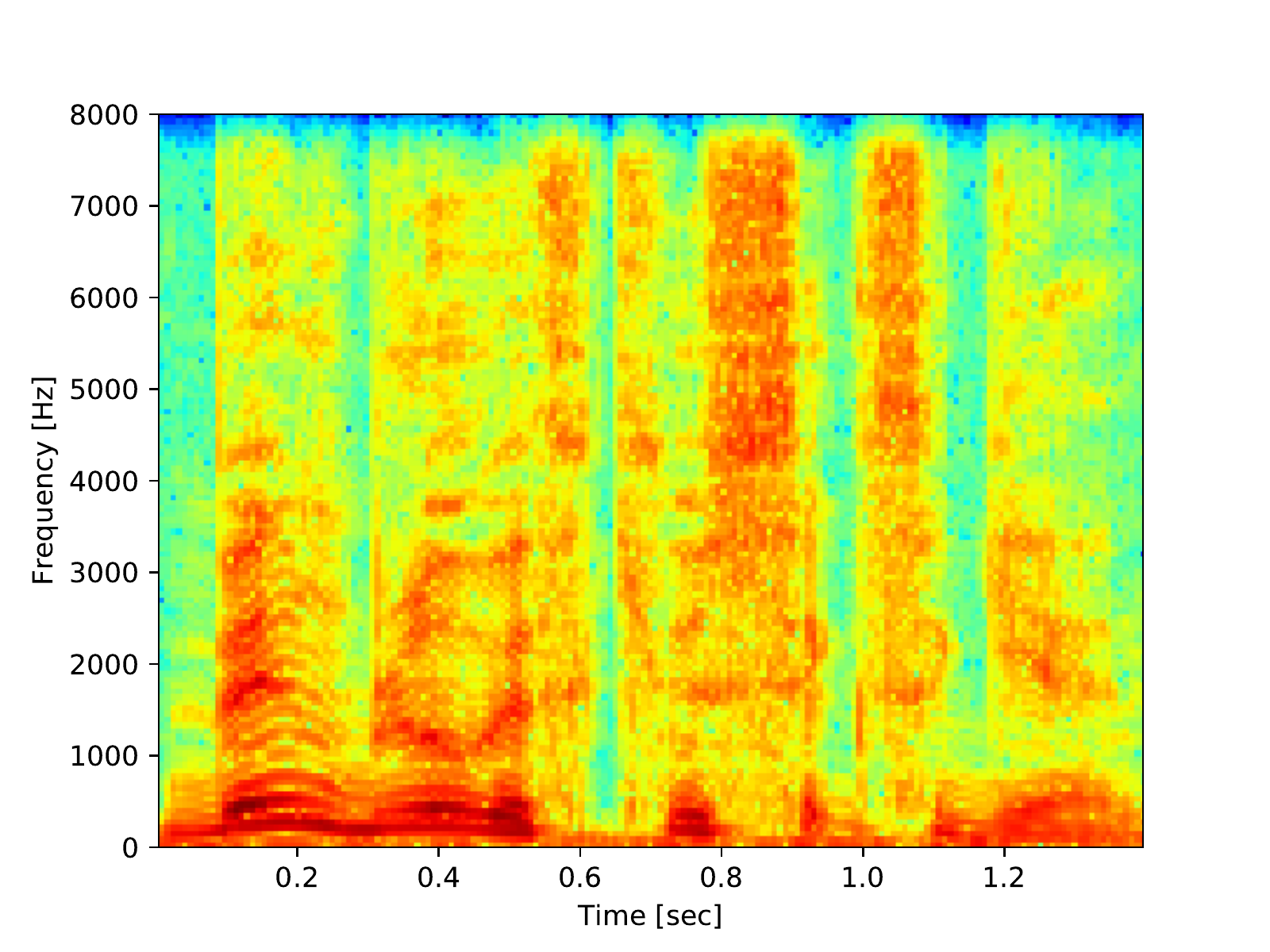}}
	\\
	\caption{Spectrogram of a  randomly enhanced -6 dB utterance from GRID + ChiME3 Speaker independent test set. It can be seen that A-only, and AV CochleaNet outperformed SS, LMMSE and SEGAN based enhancement. It is to be noted that, AV CochleaNet recovered some frequency components better than A-only CochleaNet.}
	\label{fig:spectrogramComparison}
\end{figure}

Figure~\ref{fig:spectrogramComparison} presents the noisy, clean spectrogram and spectrograms for the reconstructed speech signal of a random utterance from GRID + ChiME 3 AV corpus using SS, LMMSE, SEGAN+, A-only CochleaNet, AV CochleaNet and Oracle IBM. It is to be noted that, the speech is completely swamped with background noise and the performance of CochleaNet models can be seen (i.e.\ close to the Oracle IBM).

\subsection{Subjective testing on ASPIRE Corpus}\label{subsec:subjective-testing-on-aspire-corpus}

In the literature, significant number of objective metrics~\cite{pesq, stoi, sisdr} have been proposed to computationally approximate the subjective listening tests.
However, the only way to quantify the subjective quality is to ask listeners for their opinions.
We used MUSHRA-style~\cite{recommendation20011534} listening test method for subjective evaluation, using enhanced speech from real noisy ASPIRE corpus (section \ref{sec:aspire-corpus}).
A total of 20 native English speakers with normal-hearing participated in the listening test.
The individual test consist of 20 randomly selected utterances drawn from the ASPIRE corpus.
The first two screens were used to train participants to adjust the volume and to familiarise with the screen and the task.
In each screen, the participant were asked to score the quality of each audio sample, on a scale from
$[0, 100]$, generated by each SE model for the same sentence.
The range from $[80, 100]$ is described as “excellent”, from $[60, 80]$ as “good”, from $[40, 60]$ as “fair”, from $[20, 40]$ as “poor”, and from $[0, 20]$ as “bad”.
Noisy speech was included in the test so that participants would have a reference for the degraded speech as well as for checking if participants go through the material.

 \begin{figure}[!t]
    \centering
    \includegraphics[width=0.8\linewidth]{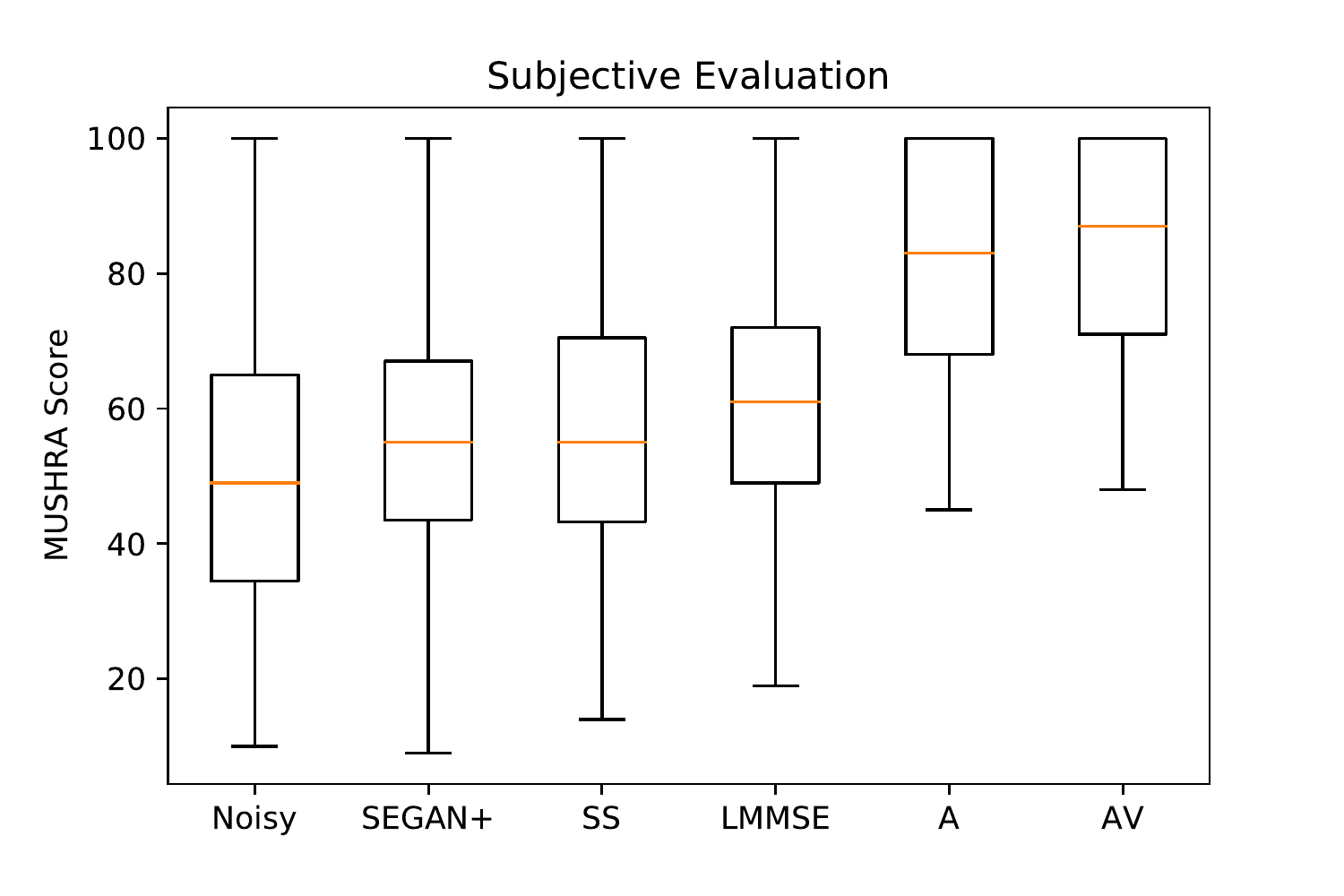}
    \caption{Result of MUSHRA listening test for ASPIRE corpus for the reconstructed speech signal using  SS~\cite{boll1979spectral}, LMMSE~\cite{ephraim1985speech}, SEGAN+~\cite{pascual2017segan}, A-only CochleaNet, AV CochleaNet.  The reference MUSHRA score for the unprocessed (Noisy) signal is included for relative comparison. }
    \label{fig:mushra_score}
\end{figure}
The times required to complete each screen were also recorded and used for removing any outliers.
We evaluated five SE models including: SEAGN, SS, LMMSE, A-only CochleaNet and AV CochleaNet.
Figure~\ref{fig:mushra_score} shows the boxplot of listeners responses in terms of the rank order of systems for the ASPIRE corpus.
The listening test results show that the superior performance of our AV CochleaNet, over A-only CochleaNet, SEGAN, spectral subtraction (SS), and log-minimum mean square error (LMMSE) based SE methods.
The results demonstrate the capability of CochleaNet to deal with the reverberation caused by multiple competing background sources observed in real-world noisy environment, by exploiting the audio and visual cues. In addition, the results show that an AV model trained on synthetic additive mixtures generalise well real noisy corpus.

\subsection{Additional Analysis}
\begin{figure}[!t]
    \centering
    \includegraphics[width=0.8\linewidth]{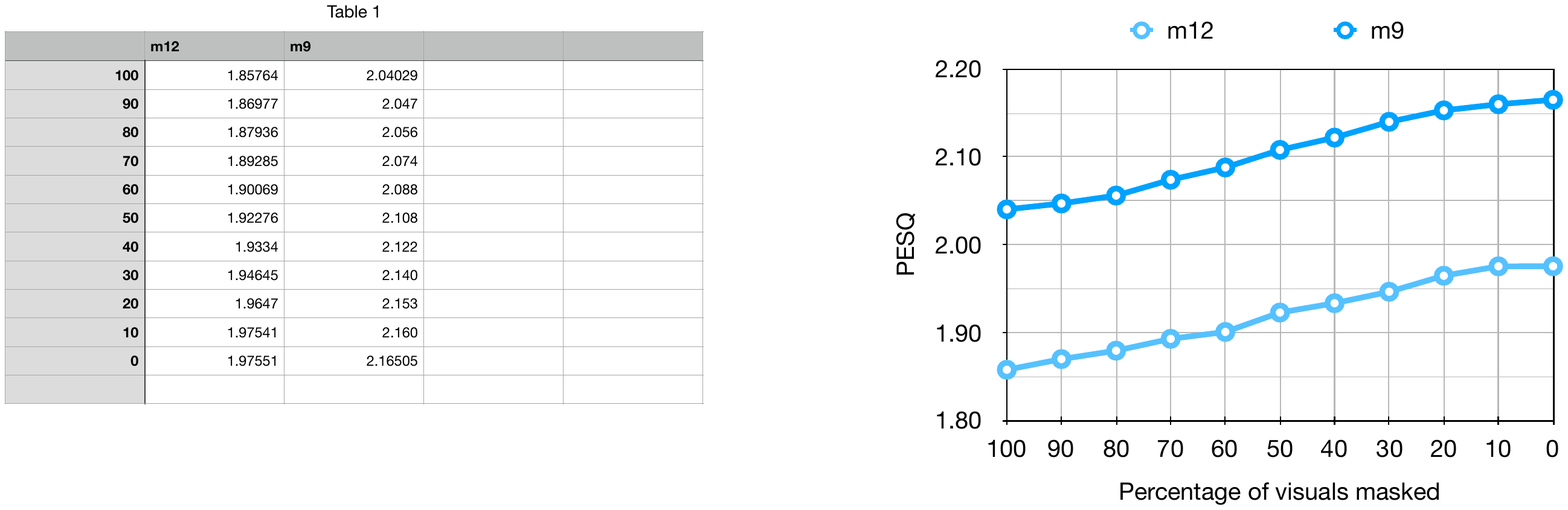}
    \caption{PESQ scores for different percentage of masked lip images}
    \label{fig:pesq_occlusion}
\end{figure}

\paragraph{Effect of occluded visual information} The model is trained and evaluated on a professionally recorded corpus that ensured none of the visual frames consists of occluded lip images (except a small number of Grid corpus utterances where visuals are absent). However, in real life scenarios specifically when the source and the target is non-stationary the model needs to be robust against the missing visual information. Therefore, to experimentally evaluate the trained AV CochleaNet behaviour in such conditions we randomly replaced a percentage of lip images with a blank visual frame. The results for lip occlusion is depicted in Figure~\ref{fig:pesq_occlusion}. It can be seen that, for both -9 dB and -12 dB, as the visual occlusion increases the PESQ score initially remains constant and after 20\% occlusion linearly starts decreasing. It is worth mentioning that, AV model performs similar to the A-only model when visuals are completely absent even though the model has not encountered such situation during training.

\paragraph{Phoneme level comparison of audio-only and audio-visual CochleaNet}
It is well known in the literature that, visual information help disambiguate the phonological ambiguity. In addition, some phonemes such as /p/ are visually distinguishable and phonemes such as /g/ cannot be visually distinguished. However, the relationship between the visually distinguishable phonemes and the AV SE performance is not known. Therefore, we conducted comparative listening tests with 3 listeners and 1000 random enhanced utterances from Grid CHiME 3 speaker independent test set to empirically identify if there is a relation between the visually distinguishable phonemes and the phonemes that AV CochleaNet can enhance better than A-only CochleaNet. The listening tests revel that AV model enhanced the /r/, /p/, /l/, /w/, /EH1/, /AE1/, /IY1/, /EY1/, /AA1/ and /OW1/ phonemes better than A-only model and the AV performance on phoneme such as /h/, /g/ and /k/ was similar to A-only performance. This confirmed the hypothesis that there is a direct relation between visually distinguishable phonemes and the phonemes that AV model works better on.

\begin{figure}[!t]
	\centering
	\subfloat[Noisy]{\includegraphics[width=0.33\textwidth]{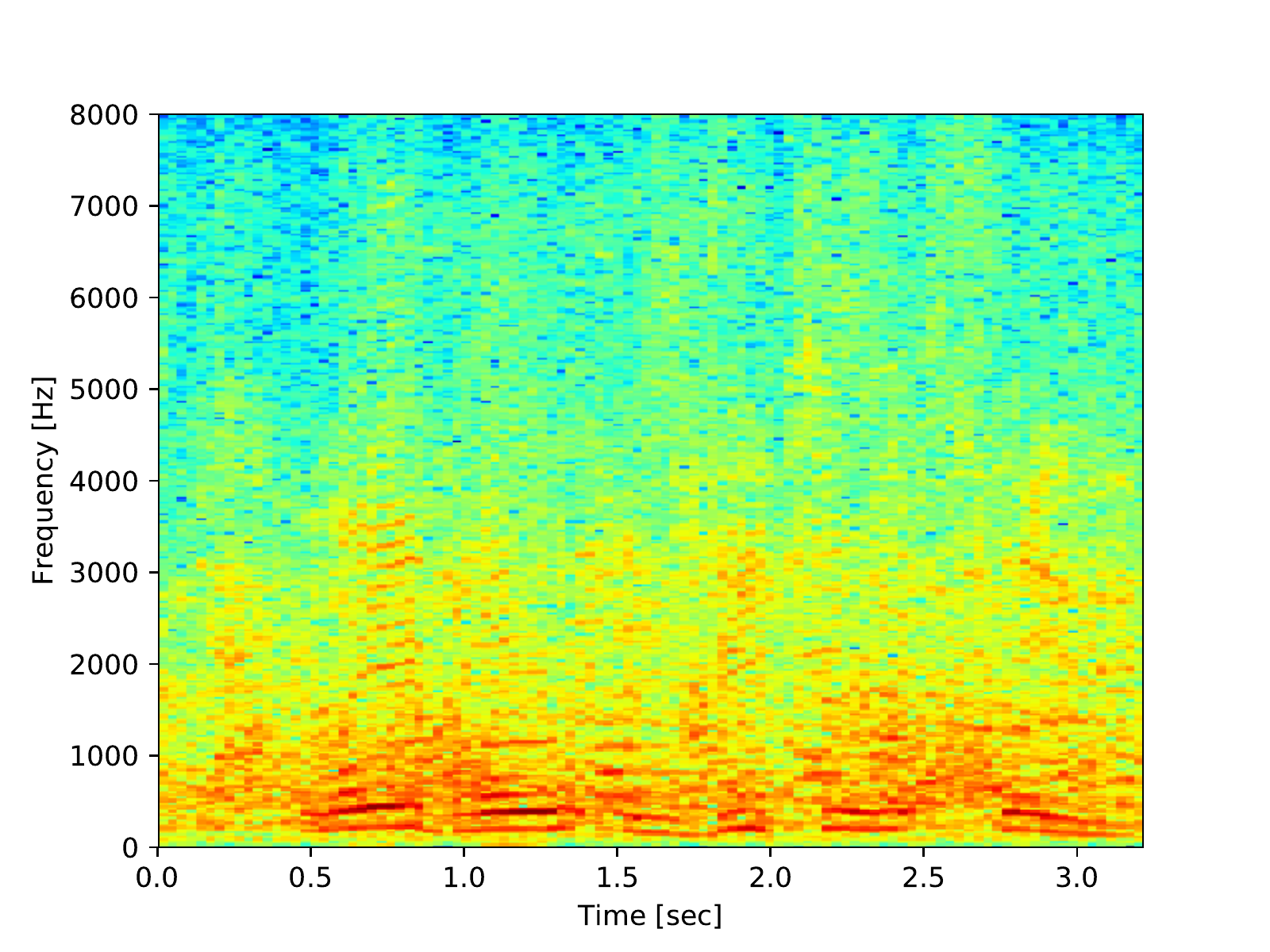}}
	\subfloat[Spectral Subtraction Enhanced]{\includegraphics[width=0.33\textwidth]{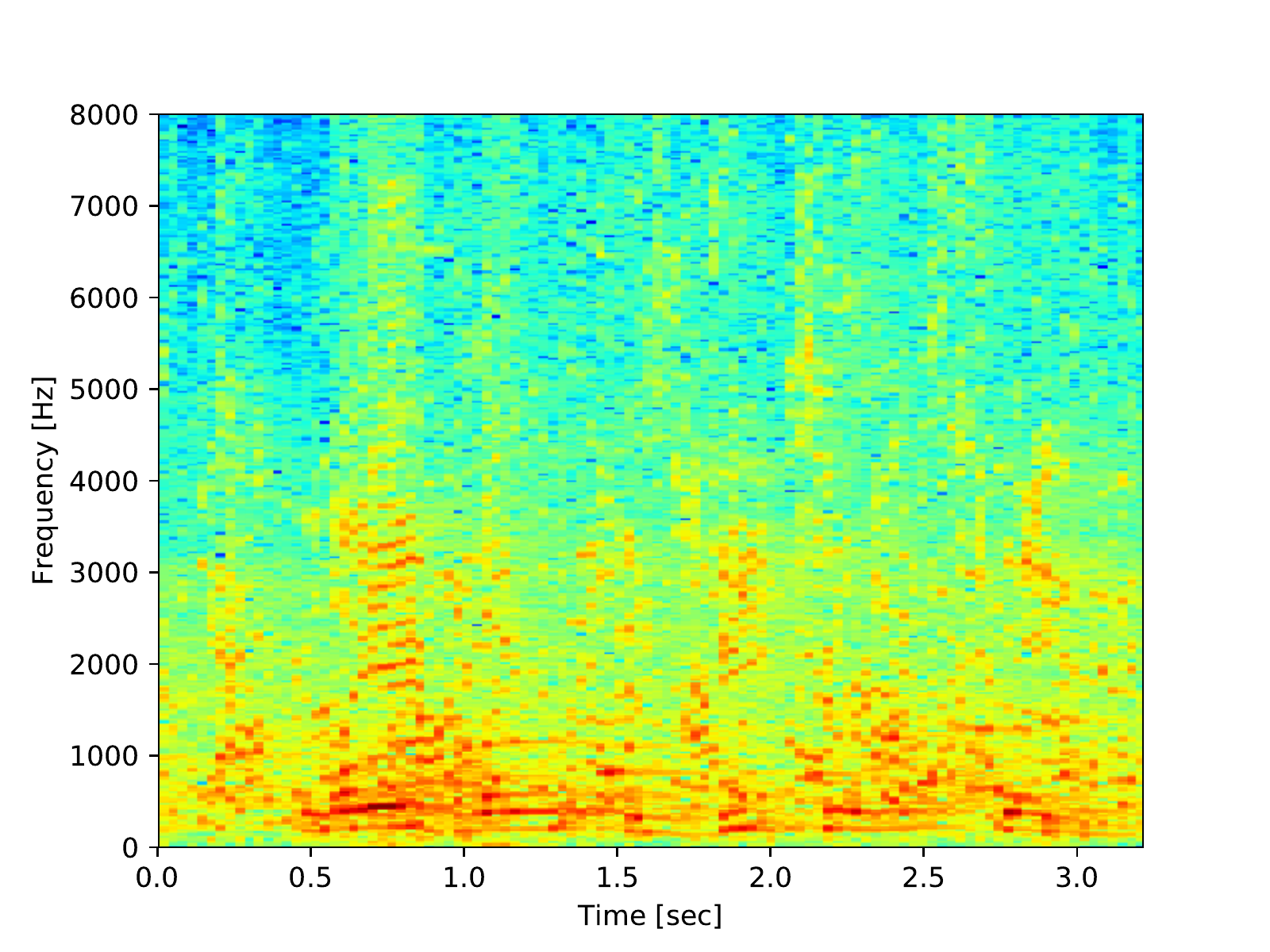}}
	\subfloat[Log MMSE Enhanced]{\includegraphics[width=0.33\textwidth]{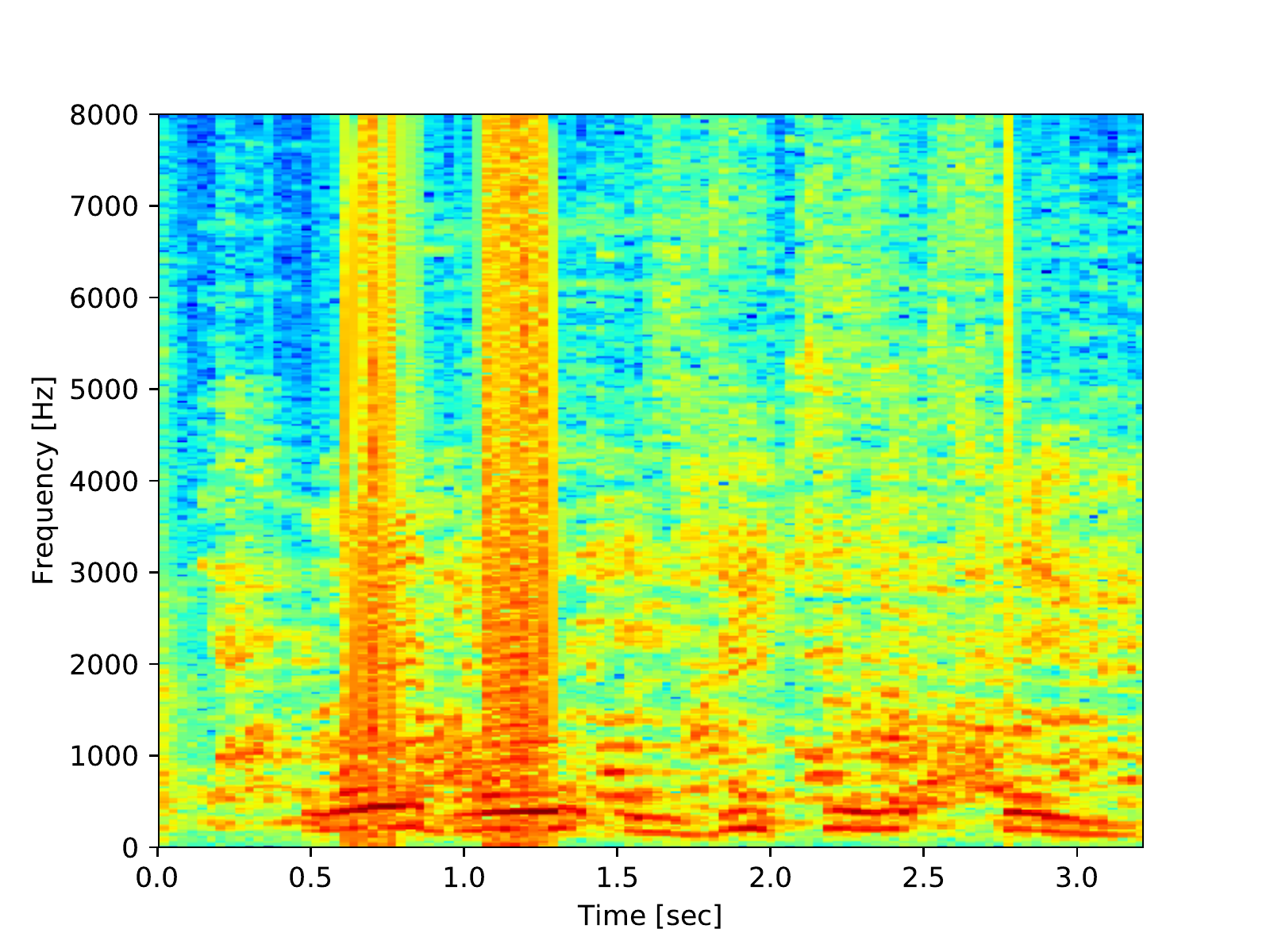}}\\
	\subfloat[SEGAN+~\cite{pascual2017segan}]{\includegraphics[width=0.33\textwidth]{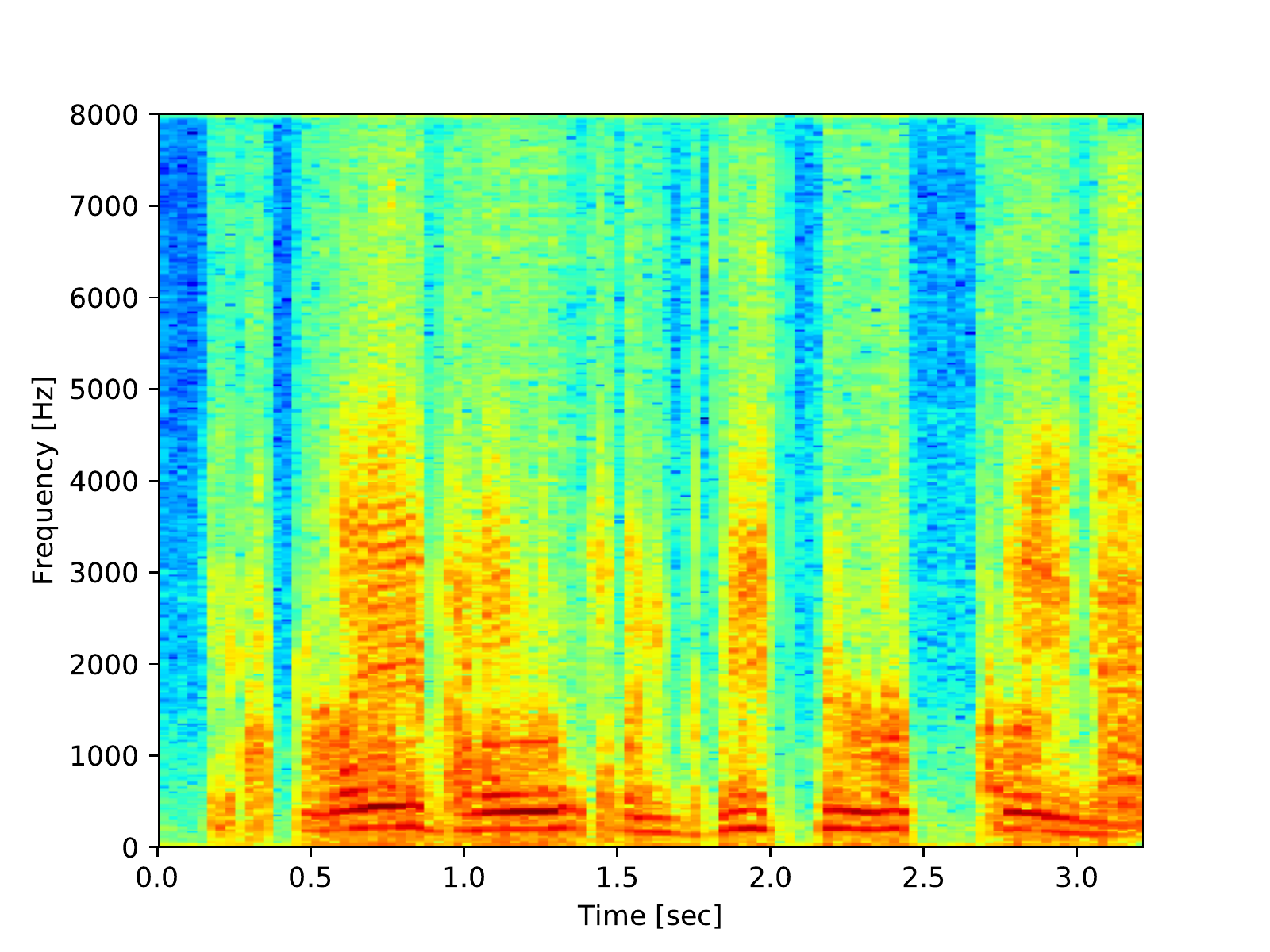}}
	\subfloat[A-only CochleaNet ]{\includegraphics[width=0.33\textwidth]{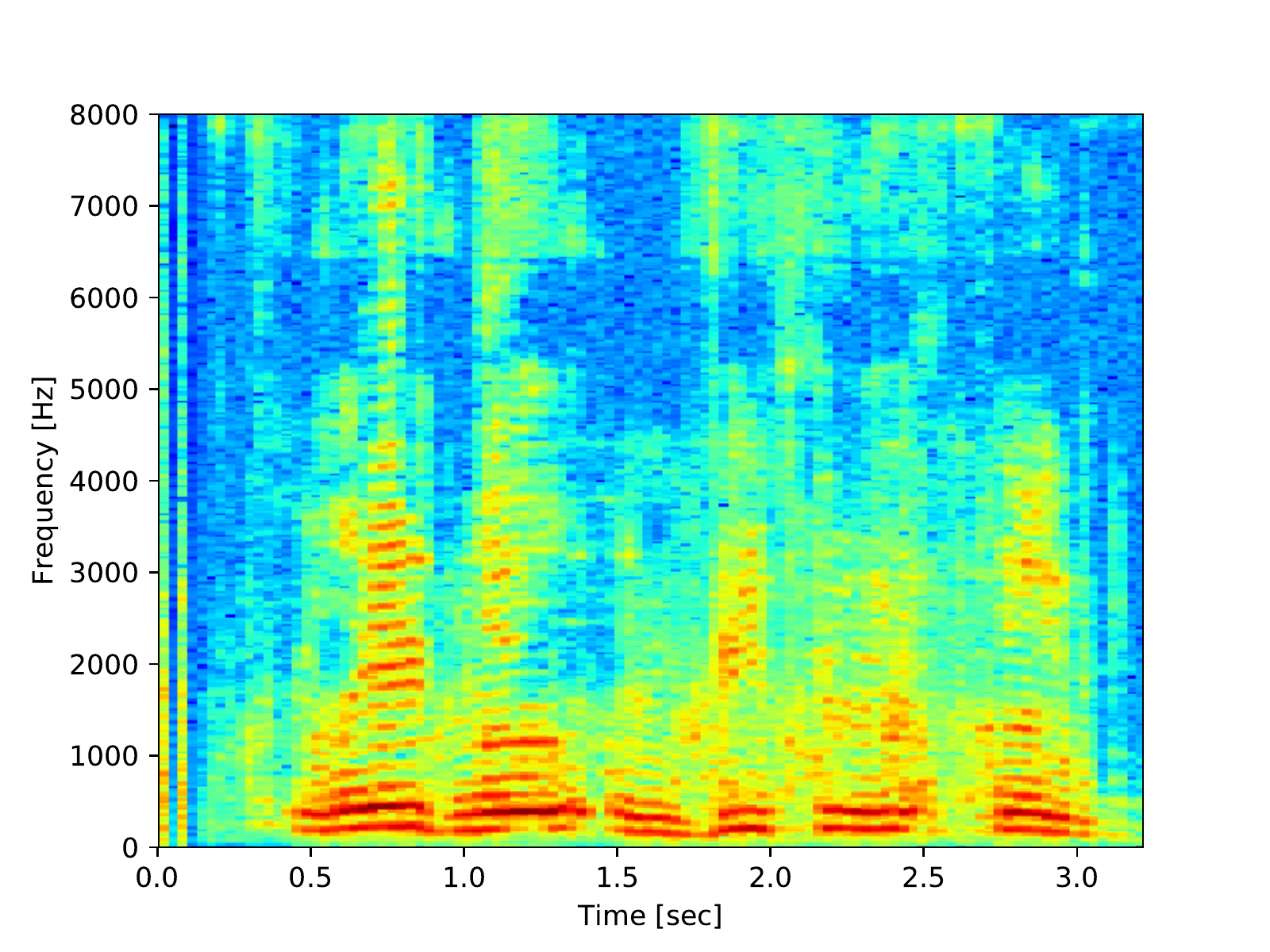}}
	\subfloat[AV CochleaNet]{\includegraphics[width=0.33\textwidth]{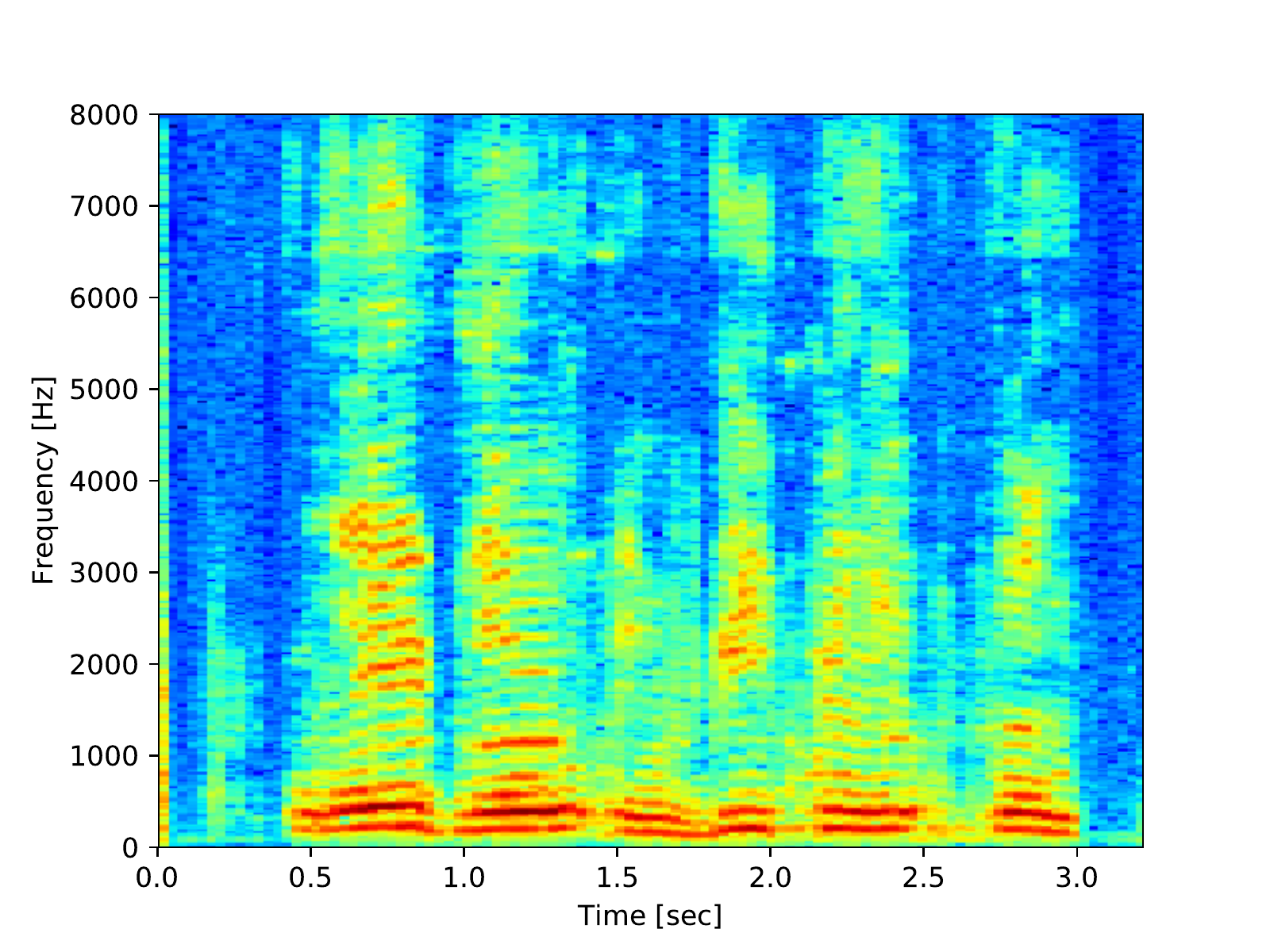}}\\
	\caption{Spectrogram of a randomly enhanced utterance from ASPIRE corpus. It is to be noted that, AV CochleaNet outperforms A-only CochleaNet, specifically in silent speech regions where visual cues (lip position) help identify if the speaker is talking or not.}
	\label{fig:spectrogramASPIREComparison}
\end{figure}

\paragraph{Comparison of audio-only and audio-visual CochleaNet in silent speech regions}
The superior performance of AV CochleaNet as compared to A-only CochleaNet could be because of the visual cues, specifically, the closed lip, could give extra information to AV model in silent speech regions. In ordered to verify this hypothesis, we calculated the mean squared error (MSE) between the predicted masks and the IBM in the silent speech regions. The A-only model achieved MSE of 0.0123 as compared to the AV that achieved MSE of 0.0108. This confirms the aforementioned hypothesis, however further analysis is needed to visualise the convolutional receptive fields and to check if a particular part of the model is active when the speaker is silent. Figure~\ref{fig:spectrogramASPIREComparison} presents the noisy spectrogram and spectrograms for the reconstructed speech signal of a random utterance from TCD-TIMIT corpus using SS, LMMSE, SEGAN+, A-only CochleaNet, AV CochleaNet. It can be seen that, the speech is completely swamped with background noise and the A-only and AV CochleaNet managed to suppress the noise dominant regions and speech dominant regions as compared to SS, LMMSE and SEGAN+. It can be seen that, in silent speech regions, AV CochleaNet outperformed A-only CochleaNet.

The main limitation with the proposed work is that: (1) the process of IBM based SE ignore the phase spectrum that lead to invalid STFT problem \cite{pandey2018new} (2) the model cannot separate the overlapping speech if more than one speaker is speaking simultaneously as the model is not trained with such mixed AV corpora (3) the ASPIRE corpus consists of only three speakers recorded in controlled real noisy environments with stationary speaker-listener setting and more challenging non-stationary real noisy corpora are required to assess the robustness of the model (4) the proposed model works only on a single channel audio and cannot exploit the binaural nature of speech we experience everyday (5) the major bottleneck in deployment of proposed mask estimation based model in listening devices such as hearing aids and cochlear implants is the data privacy concerns, high processing power requirements and processing latency.

\section{Conclusion}\label{sec:conclusion}

This paper presented a causal, language, noise and speaker independent AV DNN model for SE that contextually exploits the audio and visual cues, independent of the SNR, to estimate the spectral IBM and enhance speech.
In addition, we presented a novel AV corpus, ASPIRE\footnote{ASPIRE Corpus, enhanced speech samples, and additional supplementary material is available on the project website: \href{https://cochleanet.github.io.}{https://cochleanet.github.io}}, consisting of speech recorded in real noisy environments such as cafeteria and restaurant to evaluate the proposed model. The corpus can be used as a resource by speech community to evaluate AV SE models.
We perform extensive experiments taking into consideration the noise, speaker and language-independent criteria.
The performance evaluation in terms of objective metrics (PESQ, SI-SDR, and ESTOI) and subjective MUSHRA listening tests revealed significant improvement of our proposed AV CochleaNet as compared to the A-only CochleaNet, state-of-the-art SE (including SS, LMMSE) approaches as well as DNN based SE approaches (including SEGAN).
The simulation results have validated the phenomena of more effective visual cues at low SNRs, less effective visual cues at high SNRs.
The visual occlusion study depicts that the model performance initially remains constant till 20\% of the visuals are removed and after 20\% occlusion the performance linearly decreases as the number of occluded frame increases.
The empirical study to identify the role visual cues play in superior performance of AV model as compared to A-only model show that, there is a high correlation between visually distinguishable phonemes and the AV model performance. Moreover, the study shows that AV model significantly outperform A-only in silent speech region because it is relatively easier to audio-visually distinguish if a speaker is speaking or not as compared to only using only audio input.
In future, we intend to investigate the generalisation capability of our proposed DNN model with other more challenging conversational real noisy AV corpora.
Ongoing and future work also addresses the real time implementation challenges and privacy concerns with multimodal AV hearing aids.

\section{Acknowledgement}
This work was supported by the Edinburgh Napier University Research Studenship and UK Engineering and Physical Sciences Research Council (EPSRC) Grant No. EP/M026981/1. The authors would also like to acknowledge Dr Ricard Marxer and Prof Jon Barker from the University of Sheffield. Finally, we would like to acknowledge all the participants and support staff involved in the collection of ASPIRE corpus.

\bibliography{main.bbl}

\end{document}